\documentclass[12pt]{article}
\usepackage{amsmath,amsthm,amssymb,amsfonts,verbatim,bm,graphicx,enumerate,epstopdf,lscape,subfigure,multirow}
\usepackage{boxedminipage,longtable}
\usepackage[dvipsnames,usenames]{color}
\usepackage[pdftex,letterpaper=true,colorlinks=true,pdfpagemode=none,urlcolor=blue,linkcolor=blue, %pagebackref,
   citecolor=blue,pdfstartview=FitH,plainpages=true]{hyperref} 
\usepackage{tikz,verbatim,bibentry}
\usepackage{bm,mathrsfs,cite}
\usepackage{url} % not crucial - just used below for the URL
\usepackage[comma,authoryear]{natbib}

\newtheorem{lemma}{Lemma}
\newtheorem{theorem}{Theorem}
\newtheorem{proposition}{Proposition}

\theoremstyle{definition}
\newtheorem{remark}{Remark}
\newtheorem{example}{Example}%[section]

\newtheorem{assumption}{Assumption}

\newcommand{\utwi}[1]{}

\newcommand{\bu}{{\utwi{u}}}

\newcommand{\bA}{{\utwi{A}}}
\newcommand{\bB}{{\utwi{B}}}

\newcommand{\bD}{{\utwi{D}}}
\newcommand{\bE}{{\utwi{E}}}

\newcommand{\bI}{{\utwi{I}}}

\newcommand{\bP}{{\utwi{P}}}

\newcommand{\bU}{{\utwi{U}}}
\newcommand{\bV}{{\utwi{V}}}
\newcommand{\bW}{{\utwi{W}}}

\newcommand{\bZ}{{\utwi{Z}}}

\newcommand{\bPi}{{\utwi{\mathnormal\Pi}}}

\newcommand{\bgamma}{{\utwi{\mathnormal\gamma}}}

\newcommand{\bGamma}{{\utwi{\mathnormal\Gamma}}}
\newcommand{\bUpsilon}{{\utwi{\mathnormal\Upsilon}}}

\newcommand{\bPhi}{{\utwi{\mathnormal\Phi}}}

\newcommand{\lam}{{\lambda}}

\newcommand{\cE}{{\cal E}}

\newcommand{\cM}{{\cal M}}

\def\calM{{\cal M}}

\def\vec{\hbox{\rm vec}}

\def\Var{\textsf{var}}
\def\E{\mathbb{E}}
\def\P{\mathbb{P}}

\def\R{\mathbb{R}}

\def\lam{\lambda}
\def\real{{\mathbb{R}}}
\def\R{{\real}}

\newcommand{\bel}{\begin{eqnarray}\label}
\newcommand{\eel}{\end{eqnarray}}
\newcommand{\bes}{\begin{eqnarray*}}
\newcommand{\ees}{\end{eqnarray*}}
\newcommand{\bei}{\begin{itemize}}
\newcommand{\eei}{\end{itemize}}
\newcommand{\beiftnt}{\begin{itemize}\footnotesize}
\def\benu{\begin{enumerate}}
\def\eenu{\end{enumerate}}

\def\real{{\mathbb{R}}}
\def\R{{\real}}

\def\P{{\mathbb{P}}}

\def\complex{\mathop{{\rm I}\kern-.58em\hbox{\rm C}}\nolimits}

\def\diag{\hbox{\rm diag}}

\def\Var{\hbox{\rm var}}

\def\bA{{A}}

\def\bB{{B}}

\def\bD{{D}}

\def\etil{\widetilde{e}}

\def\bE{{E}}

\def\bI{{I}}

\def\calM{{\cal M}}

\def\bP{{P}}

\def\bPhat{\widehat{ P}}

\def\bu{{u}}

\def\bU{{U}}

\def\bV{{V}}

\def\bVhat{\widehat{ V}}

\def\bW{{W}}

\def\bWhat{\widehat{ W}}

\def\bZ{{Z}}

\def\bgamma{{\gamma}}

\def\bgammahat{\widehat{ \gamma}}

\def\bGamma{{\Gamma}}

\def\bGammahat{\widehat{ \Gamma}}

\def\eps{\epsilon}

\def\lam{\lambda}

\def\bmuhat{\widehat{ \mu}}

\def\bPi{{\Pi}}

\def\hrho{\widehat{\rho}}

\def\bSigma{{\Sigma}}

\def\bSigmahat{\widehat{ \Sigma}}
\def\bSigmatil{{\widetilde { \Sigma}}}

\def\bPhi{{\Phi}}

\def\vec{\hbox{\rm vec}}

\def\wh{\widehat}
\def\wt{\widetilde}
\newcommand{\be}{{\mathbf e}}

\newcommand{\blind}{1} 

\def\spacingset#1{\renewcommand{\baselinestretch}%
{#1}\small\normalsize} \spacingset{1.55}
\begin{document}

\if1\blind
{ 
  \title{\vspace{-0.6in} \Large \bf Simultaneous Decorrelation of Matrix Time Series
  \thanks{\footnotesize{Han was supported in part by National Science Foundation grant IIS-1741390. Chen was supported in part by National Science Foundation grants DMS-1503409, DMS-1737857 and IIS-1741390. Zhang was supported in part by NSF grants DMS-1721495, IIS-1741390 and CCF-1934924. Yao was supported in part by the U.K. Engineering and Physical Sciences Research Council grant EP/V007556/1.}} \vspace{-0.3in} }
  \author{Yuefeng Han \\
    Department of Applied and Computational Mathematics and Statistics, \\University of Notre Dame, Notre Dame, IN\\
    Rong Chen \qquad Cun-Hui Zhang \\ %\hspace{.2cm}\\
    Department of Statistics, Rutgers University, Piscataway, NJ\\
    and Qiwei Yao \\%\thanks{\footnotesize{}} \\
    Department of Statistics, London School of Economics, London, U.K.}
    \date{}
    
  \maketitle
} \fi

\if0\blind
{
  \bigskip
  \bigskip
  \bigskip
  \begin{center}
    {\LARGE\bf Simultaneous Decorrelation of Matrix \bigskip\\ Time Series}
\end{center}
  \medskip
} \fi

%\title{\LARGE \textbf{Simultaneous Decorrelation of Matrix Time Series\footnote{Han was supported in part by National Science Foundation grant IIS-1741390. Chen was supported in part by National Science Foundation grants DMS-1503409, DMS-1737857 and IIS-1741390. Zhang was supported in part by NSF grants DMS-1721495, IIS-1741390 and CCF-1934924. Yao was supported in part by the U.K. Engineering and Physical Sciences Research Council grant EP/V007556/1.}}}

%\author{Yuefeng Han \qquad Rong Chen \qquad Cun-Hui Zhang\\[.5ex]
%Department of Statistics, Rutgers University, Piscataway, NJ 08854-8019, U.S.A.\\[.5ex]
%yuefeng.han@rutgers.edu \;\; rongchen@stat.rutgers.edu \;\; czhang@stat.rutgers.edu\\[1ex]
%Qiwei Yao\\[.5ex]
%   Department of Statistics, London School of Economics, London,
%WC2A 2AE, U.K.\\
%q.yao@lse.ac.uk}
%\date{}
%\maketitle

%\vspace{0.5in}
\vspace{-0.3in}
\begin{abstract}
\vspace{-0.1in}
%\linespread{1.2}
We propose a contemporaneous bilinear transformation for a $p\times q$ matrix time series to alleviate the difficulties in modeling and forecasting matrix time series when $p$ and/or $q$ are large. %the difficulties in modelling and forecasting large number of time series together. %More precisely 
The resulting transformed matrix assumes a block structure consisting of several small matrices, and those small matrix series are uncorrelated across all times. Hence an overall parsimonious model is achieved by modelling each of those small matrix series separately without the loss of information on the linear dynamics. Such a parsimonious model often has better forecasting performance, even when the underlying true dynamics deviates from the assumed uncorrelated block structure after transformation.
% We adopt the bilinear transformation such that the rows and the columns of the matrix do not mix together, as they typically represent radically different features. %As the targeted transformation is not unique, we identify an ideal version through a new normalization, which facilitates the no-cancellation accumulation of the information from different time lags. 
The %non-asymptotic error bounds of the estimated transformation are derived, leading to the %uniform convergence rates of the estimation
uniform convergence rates of the estimated transformation are derived, which vindicate an important virtue of the proposed bilinear transformation, i.e. it is technically equivalent to the decorrelation of a vector time series of dimension max$(p,q)$ instead of $p\times q$. The proposed method is illustrated numerically via both simulated and real data examples.  
\end{abstract}

%{\sl Some key words}: 
{\it Keywords:}
Decorrelation transformation;
Eigenanalysis;
Matrix time series; 
Forecasting;
Uniform convergence rates.
\vfill

\section{Introduction}
Let $X_t= (X_{t,i,j})$ be a $p\times q$ matrix time series, i.e. there are $p\times q$ recorded values at each time from, for example, $p$ individuals and over $q$ indices or variables. Data recorded in this form are increasingly common in this information age, due to the demand to solve practical problems from, among others, signal processing, medical imaging, social networks, IT communication, genetic linkages, industry production and distribution, economic activities and financial markets. Extensive developments on statistical inference for matrix data under i.i.d. settings can be found in \cite{negahban2011estimation,rohde2011,xia2019statistical}, and the references therein. Many matrix sequences are recorded over time, exhibiting significant serial dependence which is valuable for modelling and future prediction. %Consequently there is a growing interest
The surge of development in analyzing matrix time series includes bilinear autoregressive models \citep{Hoff2015,chen2021autoregressive,xiao2021},
factor models based on Tucker's decomposition for tensors \citep{wang2019, chen2022factor, chen2020constrained, han2020, chen2020modeling,han2022rank}, and factor models based on the tensor CP decomposition \citep{han2021CP,chy2021,han2022tensor}.

A common feature in the aforementioned approaches is dimension-reduction, as the so-called `curse-of-dimensionality' is more pronounced in modelling time series than i.i.d. observations, which is exemplified by the limited practical usefulness of vector ARMA (VARMA) models. Note that an unregularized VAR(1) model with dimension $p$ involves at least $p^2$ parameters.
Hence finding an effective way to reduce the number of parameters is of fundamental importance in modelling and forecasting high dimensional time series. In the context of vector time series, most available approaches may be divided into three categories: (i) regularized VAR or VARMA methods incorporating LASSO or alternative penalties \citep{basu2015, guo2016high, lin2017, gao2019, zhou2018, ghosh2019, han2020high, han2020sparse, han2023high}, (ii) factor models of various forms \citep{pena1987identifying, bai2002, forni2005generalized, lam2012}, (iii) various versions of independent component analysis \citep{back1997first, tiao1989model, huang2014refined, matteson2011dynamic, chang2018}. Note that the literature in each of the three categories is large, it is impossible to list all the relevant references here.

In this paper we propose a new parsimonious approach for analyzing matrix time series,
%It transforms a high-dimensional dynamic modelling and/or forecasting problem into several lower-dimensional ones.
which is in the spirit of independent component analysis. It transforms a $p\times q$ matrix series into the new matrix series of the same size by a contemporaneous bilinear transformation (i.e. the values at different time lags are not mixed together). The new matrix series is divided into several submatrix series and those submatrix series are uncorrelated across all time lags. Hence an overall parsimonious model can be developed 
as those submatrix series can be modelled separately without the loss of information on the overall linear dynamics. This is particularly advantageous for forecasting future values. In general cross-serial correlations among different component series are valuable for future prediction.
However the gain from incorporating those cross-serial correlations directly in a moderately high dimensional model is typically not enough to
offset the error resulted from estimating the additional large number of parameters. This is why forecasting a large number of time series together based on a joint time series model can often be worse than that forecasting each component series separately by ignoring cross-serial correlations completely. 
%See \cite{chang2018} and also Section \ref{section:data} below.
The proposed transformation alleviates the problem by channeling all cross-serial correlations into the transformed submatrix series and those subseries are uncorrelated with each other across all time lags.  Hence the relevant information can be used more effectively in forecasting within each small models. Note that the transformation is one-to-one, therefore the good forecasting performance of the transformed series can be easily transformed back to the forecasting for the original matrix series. 
%Empirical evidence indicates that the decorrelation improves the inference efficiency. % \citep{chang2017testing}. 
Our empirical study in Section \ref{sec4} also demonstrates that, even when the underlying model does not exactly follow the assumed uncorrelated block structure, the decorrelation transformation often produces superior out-of-sample prediction, as the transformation enhances the within block autocorrelations, and the cross-correlations between the blocks are weaker or too weak to be practically useful.
%due to its overall paramounious structure with enhanced within block dynamics and weak, if any, correlation between the blocks that are ignored for model efficiency.

The basic idea of the proposed transformation is similar to the so-called principal component analysis for time series (TS-PCA) of \cite{chang2018}. Actually one might be tempted to stack a $p\times q$ matrix series into a $(pq)\times 1$ vector time series, and apply TS-PCA directly. This requires to search for a $(pq)\times (pq)$ transformation matrix. In contrast, the proposed bilinear transformation is facilitated by two matrices of size, respectively, $p\times p$ and $q\times q$. Indeed our asymptotic analysis indicates that the proposed new bilinear transformation is technically equivalent to a TS-PCA transformation with dimension max$(p,q)$ instead of $(p\times q)$; see Remark \ref{rm-rates} in Section \ref{section:theories} below. Furthermore the bilinear transformation does not mix rows and columns together, as they may represent radically different features in the data. See, e.g., the example in Section \ref{section:data} for which the bilinear transformation also outperforms the vectorized TS-PCA approach in a post-sample forecasting.

The rest of the paper is organized as follows. The targeted bilinear decorrelation transformation is characterized in Section \ref{sec2}. We introduce a new normalization under which the required transformation consists of two orthogonal transformations. The orthogonality allows us to accumulate the information from different time lags without cancellation. The estimation of the bilinear transformation boils down to the
eigenanalysis of two positive-definite matrices of sizes $p\times p$ and $q\times q$ respectively. Hence the computation can be carried out on a laptop/PC with $p$ and $q$ equal to a few thousands. Theoretical properties of the proposed estimation are presented in Section
\ref{section:theories}. The non-asymptotic error bounds of the estimated transformation are derived based on some concentration 
inequalities (e.g., \cite{rlo2000theorie,merlevede2011bernstein}); further leading to the uniform convergence rates of the estimation. 
%The consistency of the proposed latent uncorrelated blocks selection method is also presented. 
Numerical illustration with both simulated and real data is reported in Section \ref{sec4}, which shows superior performance in forecasting over the methods without transformation and TS-PCA. Once again it reinforces the fact that the cross-serial correlations are important and useful information for future prediction for a large number of time series, and, however, it is necessary to adopt the proposed decorrelation transformation (or other dimension-reduction techniques) in order to use the information effectively. All technical proofs are relegated to a supplementary.

\section{Methodology} \label{sec2}

\subsection{Decorrelation transformations} \label{sec21}

Again, let $X_t= (X_{t,i,j})$ be a $p\times q$ matrix time series.
%, i.e. there are $p\times q$ recorded values at each time from, for example, $p$ individuals and over $q$ indices or variables. 
We assume that $X_t$ is weakly stationary in the sense that all the first two moments are finite and time-invariant. Our goal is to seek for a bilinear transformation such that the transformed matrix series admits the following segmentation structure:
%the ultimate goal is to build a dynamic model for $ X_t$ and to forecast the future
%values $ X_{T+1},  X_{T+2}, \cdots.$ With moderately large $p$ and $q$,
%any direct attempts based on ARMA framework are unlikely to be successful
%due to overparametrization. % \citep{chang2018}.
%We assume that $ X_t$ admits a latent
% natural data generating process in a -- WHY NATURAL???
%segmentation structure
\begin{equation} \label{a1} %\label{b1}
 X_t = \bB \bU_t \bA^\top, \qquad
 \bU_t = \left(
\begin{array}{cccc}
\bU_{t,1,1} & \bU_{t, 1, 2}& \cdots & \bU_{t, 1, n_c} \\
\bU_{t,2,1} & \bU_{t, 2, 2}& \cdots & \bU_{t, 2, n_c} \\
\vdots & \vdots & \vdots &\vdots\\
\bU_{t,n_r,1} & \bU_{t, n_r, 2}& \cdots & \bU_{t, n_r, n_c}
\end{array}
\right),
\end{equation}
where $A$ and $B$ are unknown, respectively, $q\times q$ and $p\times p$ invertible constant matrices, and random matrix $U_t$ is unobservable,
%and is of a block structure 
%\begin{equation} \label{b1}
%\bU_t = \left(
%\begin{array}{cccc}
%\bU_{t,1,1} & \bU_{t, 1, 2}& \cdots & \bU_{t, 1, n_c} \\
%\bU_{t,2,1} & \bU_{t, 2, 2}& \cdots & \bU_{t, 2, n_c} \\
%\vdots & \vdots & \vdots &\vdots\\
%\bU_{t,n_r,1} & \bU_{t, n_r, 2}& \cdots & \bU_{t, n_r, n_c}
%\end{array}
%\right)
%\end{equation}
$U_{t, i, j}$ is a $p_i\times q_j$ matrix with unknown $p_i$ and $q_j$, $\sum_{i=1}^{n_r}p_i=p$, and $\sum_{j=1}^{n_c}q_j=q$. Furthermore,
Cov$\{ \vec(\bU_{t+\tau,i_1,j_1}), \vec(\bU_{t,i_2,j_2})\} =0$ for any $(i_1, j_1)\neq (i_2, j_2)$ and any integer $\tau$,
% $1 \leq i_1, i_2 \leq n_r$, $1 \leq j_1, j_2 \leq n_c$, 
i.e. all those submatrix series are uncorrelated with each other across all time lags.
%Our model is bilinear in the parameters, that is, linear in $A$ and linear in $B$, but not linear in $(A,B)$. 

\begin{remark} \label{rm-model}
(i) The decorrelation bilinear transformation (\ref{a1}) is in the same spirit as TS-PCA of \cite{chang2018} which transforms linearly a vector time series to a new vector time series of the same dimension but segmented into several subvector series, and those subvectors are uncorrelated across all time lags. Thus as far as the linear dynamics is concerned, one can model each of those subvector time series separately. It leads to appreciable improvement in future forecasting, as the transformed process encapsulates all the cross-serial correlations into the auto-correlations of those uncorrelated subvector processes. 
%However those segmented subvector series are not guaranteed to exist. This is a marked difference from the standard PCA. Nevertheless even then the TS-PCA transformation leads to some approximate and practically beneficial segmentations which ignore some weak correlations of little value. 
%The proposed procedure is an extension of TS-PCA to matrix time series, with several important distinctions.

(ii) In the matrix time series setting, one would be tempted to stack all the elements of $ X_t$ into a long vector and to apply TS-PCA of \cite{chang2018} directly, though it destroys the original matrix structure.
%\rc{and loses the important column and row classification information}. 
This requires to search for the $(pq) \times (pq)$ decorrelation transformation matrix $\Phi$ such that $\vec( X_t) = \Phi \vec(U_t)$. The proposed model (\ref{a1}) is to impose a low-dimensional structure $\Phi=A\otimes B$, where $\otimes$ denotes the matrix Kronecker product, such that the technical difficulty is reduced to that of estimating two transformation matrices of size $p\times p$ and $q\times q$ respectively, with significant faster convergence rate; see Remark \ref{rm-rates} in Section \ref{section:theories} below.
The empirical results in Section \ref{sec4} also demonstrate the benefit of maintaining the matrix structure in out-sample prediction performance.
%Also note that simpler models are more likely to be applicable to smaller
%matrix series. For example, the bilinear AR models of
%\citet{chen2021autoregressive} with small
%orders are more plausible for $\bU_{t,i,j}$ than for $ X_t$, the factor models of
%\cite{wang2019} or \cite{chy2021} with small number of factors are more 
%appropriate for $\bU_{t,i,j}$ than for $X_t$.
  
(iii) The segmentation structure in (\ref{a1}) may be too rigid and the division of the submatrices may not be that regular. In practice a small number of row blocks and/or column blocks may be resulted from applying the estimation method proposed in Section \ref{section:estimation} below. Then applying the same method again to each submatrix series $U_{t, i,j}$ may lead to a finer segmentation with irregularly sized small blocks. %\yfm{For example, \rc{if the underlying true $U_t$ has the structure} $U_t=[U_{t,1}\ U_{t,2};\ U_{t,3}\ U_{t,4}]\in\R^{4\times 5}$, where $U_{t,1}\in\R^{2 \times 3}, U_{t,2}\in\R^{ 2\times 2}, U_{t,3} \in\R^{ 2 \times 1}, U_{t,4} \in\R^{2 \times 4}$. Then, the first pass \rc{of applying the proposed segmentation method would} lead to the segmentation of two submatrices of the same size $2\times 5$. Applying the segmentation method %in Section \ref{section:estimation} 
%again to each of those two submatrices separately \rc{would} lead to the correct separation of four {\rc irregular} blocks \rc{of different sizes}.}

(iv) Similar to TS-PCA, the desired segmentation structure (\ref{a1}) may not exist {in real applications}. Then the estimated bilinear transformation, by the method in Section \ref{section:estimation}, leads to an approximate segmentation which encapsulates most significant correlations across different component series into the segmented submatrices while ignoring some small correlations. With enhanced auto-correlations, those submatrix processes become more predictable while those ignored small correlations are typically practically negligible. Consequently the improvement in future prediction
still prevails in spite of the lack of an exact segmentation structure. See \cite{chang2018}, and also a real data example in Section \ref{section:data} below in which the three versions of segmentation (therefore, at least two of them are approximations) uniformly outperform the various methods without the transformation. A simulation study in Section~\ref{section:simulation} also shows that when the underlying true model deviates from the desired segmentation structure by a moderate amount, the approximated segmentation model produces superior out-sample prediction performance than those without the transformation.
%, as these models suffer from large estimation errors in estimating a large number of parameters.

(v) The use of bilinear form in \eqref{a1} is common in matrix and tensor data analysis. For example, it is used in regression with matrix-type covariates \citep{basu2012blr,zhou2013tensor,wang2019symmetric}, matrix autoregressive model \citep{chen2021autoregressive,Hoff2015} and matrix factor model \citep{wang2019, chen2020constrained, chen2022factor}.
%, in which such bilinear structure is also adopted. 
%\rc{Interpretation of the parameters in $A$ and $B$ is facilitated by noting that the rows and the columns of $X_t$ are linear combinations of, respectively, the rows and the columns of $U_t$.}
%\rc{For example, if $B=I$, then each column of $X_t$ ($X_t=AU_t$) is a transformation of the corresponding column of $U_t$. In this case, the column grouping of $U_t$ would be preserved in $X_t$. \rc{In addition, as both $A$ and $B$ are invertible, $U_t=A^{-1}X_tB^{-1}$. If $B=I$, then each column of $U_t$ is obtained by the linear transformation of the corresponding column of $X_t$, using $A^{-1}$. Hence $A$ can be viewed as column transformation matrix.}}
%{\color{brown}
%I AM NOT SURE IF THIS INTERPRETATION REALLY HELPS -- QY}

%This model \rc{is} %appears similar to matrix autoregressive model \citep{chen2021autoregressive,Hoff2015} and matrix factor model \citep{wang2019}, in which such \rc{a} bilinear structure is also adopted. Interpretation of the parameters in $A$ and $B$ is facilitated by noting that} the rows and the columns of \yfm{$X_t$} are linear combinations of, respectively, the rows and the columns of \yfm{$U_t$, i.e. for a given ordered pair of nodes $(i_1,j_1)$,
%\begin{align*}
%X_{t,i_1,j_1}=\sum_{i_2=1}^p \sum_{j_2=1}^q  a_{j_1,j_2} b_{i_1,i_2} U_{t,i_2 j_2}.  
%\end{align*}}
%More interesting interpretations on integrating the column and row interactions were explicitly illustrated in \cite{wang2019}.

(vi) For dealing with time series with structural changes, it is possible to allow $A$ and $B$ to be time varying, which is technically more demanding and will be explored elsewhere.
\end{remark}

\subsection{Normalization and identification} \label{22} 

To simplify the statements, let $\E( X_t) = 0$. This amounts to centre the data by the sample mean first, which will not affect the asymptotic properties of the estimation under the stationarity assumption. Thus $\E(\bU_t) = 0$.

The terms $\bA, \bB$ and $\bU_t$ on the RHS of (\ref{a1}) are not uniquely defined, and any $\bA, \bB$ and $\bU_t$ satisfying the required condition will serve the decorrelation purpose. Therefore we can take the advantage from this lack of uniqueness to identify the `ideal' $\bA$ and $\bB$ to improve the estimation effectiveness. More precisely, by applying a new normalization, we identify $\bA$ and $\bB$ to be orthogonal, which facilitates the accumulation of the information from different time lags without cancellation. To this end, we first list some basic facts as follows.
% \bSigma_1^{(x)} = \E( X_t^\top  X_t)/p, \qquad\quad \;\;\;\; \bSigma_2^{(x)} = \E(  X_t  X_t^\top)/q,
\begin{enumerate}[(i)]
\item
$(B, U_t, A^\top)$ in (\ref{a1}) can be replaced by $(B D_1, D_1^{-1} U_t D_2^{-1}, D_2 A^\top)$ for any invertible block diagonal matrices $D_1$ and $D_2$ with the block sizes, respectively, $(p_1, \cdots, p_{n_r})$ and $(q_1, \cdots, q_{n_c})$.
%\rc{In particular, each sub-diagonal block of $D_1$ and $D_2$ can be a matrix that permutes the indices within the block.}
%
%\item \rc{$(\bB, \bU_t, \bA^\top)$ in (\ref{a1}) can be replaced by $(\bB \bD_1, \bD_1^{-1} \bU_t \bD_2^{-1}, 
In particular, $D_1$ and $D_2$ can be the permutation matrices which permute the columns and rows within each block. In addition, $D_1$ and $D_2$ can be the block matrices that permute $n_r$ row blocks and $n_c$ column blocks respectively. Note both $(p_1, \cdots, p_{n_r})$ and $(q_1, \cdots, q_{n_c})$ are defined by (\ref{a1}) only up to any permutation.

\item
The $q$ columns of $B U_t$ can be divided into $n_c$ uncorrelated blocks of sizes $(q_1, \cdots, q_{n_c})$, i.e. the columns of $\bB \bU_t$ resemble the same pattern of the uncorrelated blocks as those of $\bU_t$. Thus $\bSigma_1^{(u)} \equiv \E(U_t^\top B^\top B U_t)/p$ is a block diagonal matrix with the block sizes $(q_1, \cdots, q_{n_c})$. In the same vein, $\bSigma_2^{(u)} = \E(U_t A^\top A U_t^\top)/q$ is a block diagonal matrix with the block sizes $(p_1, \cdots, p_{n_r})$.

\item Put $B=(B_1, \cdots, B_{n_r})$ and $A= (A_1, \cdots, A_{n_c})$, where $B_i$ is $p\times p_i$ and $A_i$ is $q\times  q_i$. Then linear spaces $\calM(B_1), \cdots, \calM(B_{n_r})$ and $\calM(A_1), \cdots, \calM(A_{n_c})$ are uniquely defined by (\ref{a1}) up to any permutation, though $B$ and $A$ are not, where $\calM(G)$ denotes the linear space spanned by the columns of matrix $G$. 
Note that for any $q\times q$ positive definite matrix $\Sigma_1$ and $p\times p$ positive definite matrix $\Sigma_2$,
\[
\bSigma_1^{-1/2} \bA = (\bSigma_1^{-1/2} \bA_1, \cdots, \bSigma_1^{-1/2} \bA_{n_c}), \qquad \calM(\bA_i) =
\bSigma_1^{1/2}\calM(\bSigma_1^{-1/2} \bA_i) \;\;\; i=1, \cdots, n_c,
\]
\[
\bSigma_2^{-1/2} \bB = (\bSigma_2^{-1/2} \bB_1, \cdots, \bSigma_2^{-1/2} \bB_{n_r}), \qquad \calM(\bB_i) =
\bSigma_2^{1/2}\calM(\bSigma_2^{-1/2} \bB_i) \;\;\; i=1, \cdots, n_r.
\]
\end{enumerate}

Put 
$\bSigma_1^{(x)} = \E( X_t^\top X_t)/p$ and $\bSigma_2^{(x)} = \E( X_t X_t^\top)/q$. Proposition \ref{prop1} below indicates that if we replace $ X_t$ in (\ref{a1}) by a normalized version $\{ \bSigma_2^{(x)}\}^{-1/2}  X_t \{ \bSigma_1^{(x)}\}^{-1/2}$, we can treat both $\bA$ and $\bB$ in (\ref{a1}) as orthogonal matrices. This orthogonality plays a key role in combining together the information from different time lags in our estimation (see the proof of Proposition \ref{prop2} in the appendix).

\begin{proposition} \label{prop1}
Let both $\bSigma_1^{(x)}$ and $\bSigma_2^{(x)}$ be invertible. Then it holds that
%model (\ref{a1}) can be written as 
\begin{equation} \label{an1}
 X_t = \{ \bSigma_2^{(x)}\}^{1/2} \bB_* \bU_t^* \bA_*^\top \{ \bSigma_1^{(x)}\}^{1/2}, 
\end{equation}
where $ \bU_t^*$ admits the same segmentation structure as $\bU_t$ in (\ref{a1}), and $\bA_*$ and $ \bB_*$ are, respectively, $q\times q$ and $p\times p$ orthogonal matrices. More precisely,
\begin{equation} \label{an2}
\bU_t^* = \{ \bSigma_2^{(u)}\}^{-1/2} \bU_t \{ \bSigma_1^{(u)}\}^{-1/2}, %\;\;
\bA_* = \{ \bSigma_1^{(x)}\}^{-1/2} \bA \{ \bSigma_1^{(u)}\}^{1/2}, %\;\;
\bB_* =\{ \bSigma_2^{(x)}\}^{-1/2} \bB \{ \bSigma_2^{(u)}\}^{1/2}.
\end{equation}
\end{proposition}

The proof of Proposition \ref{prop1} is almost trivial. First note that both $\bSigma_1^{(u)}, \, \bSigma_2^{(u)}$ are invertible. Then (\ref{an1}) follows from (\ref{a1}) and (\ref{an2}) directly. The orthogonality of, e.g., $\bB_*$ follows from equality $\E[\{ \bSigma_2^{(x)}\}^{-1/2}  X_t  X_t^\top \{ \bSigma_2^{(x)}\}^{-1/2}]/q=\bI_p$, (\ref{an1}) and (\ref{an2}). Also note that $\{ \bSigma_2^{(u)}\}^{-1/2}$ and $\{ \bSigma_1^{(u)}\}^{-1/2}$ are of the same block diagonal structure as, respectively, $\bSigma_2^{(u)}$ and $\bSigma_1^{(u)}$. This implies that $\bU_t^*$ admits the same segmentation structure as $\bU_t$. 

%\begin{equation}
%\label{eq:normalize}
%\overrightarrow{ X_t}=  X_t \big(\bSigma_1^{(x)}\big)^{-1/2} \mbox{\ \ and \ \ }
%\overleftarrow{ X_t}= \big(\bSigma_2^{(x)}\big)^{-1/2} X_t 
%\end{equation} 

Write $A_* = (A_{*1}, \cdots,A_{*n_c}), B_* = (B_{*1},\cdots,B_{*n_r})$, where $A_{*j}$ has $q_j$ columns and $B_{*i}$ has $p_i$ columns. Then $U_{t,i,j}^*=A_{*j} \{ \Sigma_2^{(x)}\}^{-1/2} X_t \{ \bSigma_1^{(x)}\}^{-1/2} B_{*i}$.
Note that $(\bB_*, \bU_t^*, \bA_*^\top)$ in (\ref{an1}) are (still) not uniquely defined, similar to the property (i) above.
%as they can be replaced by $(\bB_* \bD_1, \bD_1^\top\bU_t^* \bD_2^\top, \bD_2\bA_*^\top)$ for any orthogonal block diagonal matrices $\bD_1$ and $\bD_2$ with appropriate block sizes.
In fact, only the linear spaces $\calM(B_{*1}), \cdots, \calM(B_{*n_r})$ and $\calM(A_{*1}), \cdots, \calM(A_{*n_c})$ are uniquely defined.
Proposition \ref{prop2} below shows that we can take the orthonormal eigenvectors of two properly defined positive definite matrices as the columns of $\bA_*$ and $\bB_*$. With $\bA_*$ and $\bB_*$ specified, the segmented $\bU_t^*$ can be solved from (\ref{an1}) directly. Let
% Let the auto-cross-covariance matrices between the $i$-th and the $j$-th
% rows of $\overrightarrow{ X_t}$ at lag $\tau$ be 
\bes
\bV^{(1)}_{\tau,i,j}
= \{ \bSigma_1^{(x)}\}^{-1/2}  \E({ X}_{t+\tau}^\top \bE_{i,j}{ X_t})
\{ \bSigma_1^{(x)}\}^{-1/2}, \quad
\bV^{(2)}_{\tau,i,j}
= \{ \bSigma_2^{(x)}\}^{-1/2}  \E({ X}_{t+\tau} \bE_{i,j}{ X_t}^\top)
\{ \bSigma_2^{(x)}\}^{-1/2},
\ees
where $\bE_{i,j} $ is the unit matrix with 1 at position $(i,j)$ and 0 elsewhere, and $\bE_{i,j} $ is $p\times p$ in the first equation, and $q\times q$ in the second equation. For a prespecified integer $\tau_0 \ge 1$, let
\begin{equation} \label{aa9}
\bW^{(1)} = \sum_{\tau=-\tau_0}^{\tau_0} \sum_{i=1}^p\sum_{j=1}^p
\frac{\bV^{(1)}_{\tau,i,j}\big(\bV^{(1)}_{\tau,i,j}\big)^\top}{p^2}, \qquad
\bW^{(2)} = \sum_{\tau=-\tau_0}^{\tau_0} \sum_{i=1}^q\sum_{j=1}^q
\frac{\bV^{(2)}_{\tau,i,j}\big(\bV^{(2)}_{\tau,i,j}\big)^\top}{q^2}.
\end{equation}

\begin{proposition} \label{prop2}
Let both $\bSigma_1^{(x)}$ and $\bSigma_2^{(x)}$ be invertible, and all the eigenvalues of $\bW^{(i)}$ be distinct, $i=1, 2$. Also let $\tau_0\ge1$. Then the $q$ orthonormal eigenvectors of $\bW^{(1)}$ can be taken as the columns of $\bA_*$, and the $p$ orthonormal eigenvectors of $\bW^{(2)}$ can be taken as the columns of $\bB_*$.
\end{proposition}

%\rc{Proposition \ref{prop2} can be seen as follows:} Let $W^{(1)}_z$ be the matrix defined \rc{in the same manner as $W^{(1)}$ in \eqref{aa9}}, but with the $X_t$ appeared in $V_{\tau, i,j}^{(1)}$ replaced by \yfm{$B U_t\{\Sigma_1^{(u)}\}^{-1/2}$ (i.e. $\{ \Sigma_2^{(x)}\}^{1/2} B_* U_t^*$)}. \rc{Then, it is easily shown} that $W^{(1)}_z$ is a block diagonal matrix \rc{using the fact that Cov$\{ \vec(U_{t+\tau,i_1,j_1}),\vec(U_{t,i_2,j_2})\} =0$ for any $(i_1, j_1)\neq (i_2, j_2)$ and any integer $\tau$. Since} $W^{(1)} = A_* W^{(1)}_z A_*^\top$, the orthonormal eigenvectors of $W^{(1)}$ \rc{forms a representative of $A_*$.}
%are the columns of $A_* D$, where $D$ is an orthogonal block diagonal matrix. 

The proposition above does not give any indication on how to arrange the columns of $\bA_*$ and $\bB_*$, which should be ordered according to the latent uncorrelated block structure (\ref{a1}). We address this issue in Section \ref{section:estimation} below.

\begin{remark}\label{rmk:identification}
The representation \eqref{a1} suffers from the indeterminacy due to the fact that the column blocks and row blocks can be permuted, and the columns and rows within each block can be rotated, see property (i) above. However this indeterminacy does not have material impact on identifying the desired structure (\ref{a1}), as any one of such representations will serve the purpose. The developed asymptotic theory guarantees that the proposed estimator converges to a representation which fulfills the conditions imposed on (\ref{a1}).
\end{remark}

\begin{remark}\label{rmk:trans}
In case that $W^{(1)}, W^{(2)}$ have tied eigenvalues and the corresponding eigenvectors across different blocks, Proposition \ref{prop2} no longer holds. Then the blocks sharing the tied eigenvalues cannot be separated. To avoid the tied eigenvalues, we may use different values of $\tau_0$, or different form of $W^{(i)}$. For example, we may replace $W^{(1)}$ by
\bes
W^{(1,f)} = \sum_{\tau=-\tau_0}^{\tau_0} \sum_{i=1}^p\sum_{j=1}^p
\frac{f\left( V^{(1)}_{\tau,i,j} V^{(1)\top}_{\tau,i,j} \right)}{p^2}.
\ees
where $f(V)$ is a function of a symmetric matrix $V$ in which $f(V)=\Gamma D^{(f)}\Gamma^\top$, $V=\Gamma D\Gamma^\top$ is the eigen-decomposition for $V$, and $D^{(f)}=\diag(f(d_1),\ldots,f(d_q))$ is the diagonal-element-wise transformation of the diagonal matrix $D$ of the eigenvalues.

In fact the condition that all eigenvalues of $W^{(i)}$ are different can be relaxed. For example, we only require any two eigenvalues of $W^{(i)}$ corresponding two different blocks to be different while the eigenvalues corresponding to the same black can be the same. See also a similar condition in the eigen-gap $\Delta$ in \eqref{eigen-gap} in Section \ref{section:theories} below.

% Note that these variations 
% reserve the block structure, with difference sets of eigenvalues of resulting $\bW^{(x)}$.
% A suitably chosen $\tau_0$ or the transformation $f$ can ensure that all 
% eigenvalues are different. Of course in practice the estimation errors may mask 
% the case of equal eigenvalues. More discussions of the issue will be present later. 
%\bes
%\bW^{(x,f)} = \sum_{\tau=0}^{\tau_0} \sum_{i=1}^p\sum_{j=1}^p
%\frac{f\big(\bV^{(x)}_{\tau,i,j},\big(\bV^{(x)_{\tau,i,j}\big)^\top\big)}{p^2}.
%\ees
%where $f(\bV,\bV^\top)$ be a polynomial with two $q\times q$ matrix inputs $\bV$ and $\bY$ such that $f(\bV,\bV^\top)$ is always nonnegative definite.
%Then $\bW^{(x)}$ defined in (\ref{a9}) corresponds to $f(\bV,\bY)= \bV\bY$.
%A simple alternative is $f(\bV,\bY)=\bY \bV$.
%Instead of maximizing $n_c$, we can also use the intersection of two
%nonidentical partitions, which is a finer partition.
\end{remark}

% \begin{remark}
% It is our objective to segment the matrix time series into as many uncorrelated blocks as possible (maximum segmentation), so each block is small, with a small model and less number of parameters used to fit it. However, even if some of the partition groups are (mistakenly) combined, the segmentation procedure is still useful, just not as efficient and parsimous as that with the maximum segmentation. 
% \end{remark}

% \begin{remark} One may entertain some finer partitions than (\ref{b1}).
% For example, after the initial column decorrelation
% (\ref{a11}),  one can perform row decorrelation for each
% sub $\bZ_{t,j}$ separately, resulting in a irregular tiling type of segmentation of the matrix.
% \end{remark}

%As I remembered, in the previous version (3th or 2th?), both (23) and (27) were placed in Section 2 and 3. Then, in the 4th version, they were moved to Section 4. I think it is better to move them to the end of this section 2. In the current version, we only briefly discuss the issue of tied eigenvalues in the end of this section.

\begin{remark}\label{rmk:tau0}
In practice, we need to specify $\tau_0$ in (\ref{aa9}). In principle any $\tau_0\geq 1$ can be used for the purpose of segmentation. A larger $\tau_0$ may capture more lag-dependence, but may also risk adding more `noise' when the dependence decays fast. Since autocorrelation is typically at its strongest at small lags, a relatively small $\tau_0$, such as $\tau_0\le 5$, is often sufficient \citep{lam2011,chang2018,wang2019,chen2022factor}.
\end{remark}

% Let $f(V,U)$ be a polynomial
% with two $q\times q$ matrix inputs $V$ and $U$ such that
% $f(V,V^\top)$ is always nonnegative definite. Define
% \bes
% W^{(x,f)} = \sum_{\tau=0}^{\tau_0} \sum_{i=1}^p\sum_{j=1}^p
% \frac{f\big(V^{(z,\tau,i,j)},\big(V^{(z,\tau,i,j)}\big)^\top\big)}{(\tau_0+1) p^2}
% %W^{(z,f)} &=& I_{q\times q}+\sum_{i=1}^p\sum_{j\neq i}
% %\frac{f\big(V^{(z,0,i,j)},\big(V^{(z,0,i,j)}\big)^\top\big)}{p(p-1)}
% %\cr && + \sum_{\tau=1}^{\tau_0} \sum_{i=1}^p\sum_{j=1}^p
% %\frac{f\big(V^{(z,\tau,i,j)},\big(V^{(z,\tau,i,j)}\big)^\top\big)}{\tau_0 p^2}
% \ees
% Then, $W^{(z,f)}$ has the decomposition
% \bes
% W^{(z,f)}
% = \begin{pmatrix} W^{(z,f)}_{G_1\times G_1} & 0 & \cdots & 0 \cr
% 0 & W^{(z,f)}_{G_2\times G_2} & \cdots & 0 \cr \vdots & \vdots & \cdots & \vdots \cr
% 0 & 0 & \cdots & W^{(z,f)}_{G_{n_c}\times G_{n_c}}\end{pmatrix}.
% \ees
% It is clear that $W^{(z,f)}=W^{(z)}$ when $f(V,U)=VU$. The simplest alternative to $f(V,U)=VU$ is
% $f(V,U)=UV$. Similar to $W^{(z)}$, $W^{(z,f)}$ also provides a partition of $\{1,\ldots,n_c\}$.
% The point here is that the intersection of two nonidentical partitions gives a finer partition.
% As different $f$ could be used, this process could hopefully lead to the full identification of
% the original $G_i$.

%\section{Detection of uncorrelated components}\label{section:detection}

\subsection{Estimation} \label{section:estimation}

The estimation for $\bA_*$ and $\bB_*$, defined in \eqref{an2}, is based on the eigenanalysis of the sample versions of matrices defined in (\ref{aa9}). To this end, let
\begin{equation} \label{est-0}
\wh{\bSigma}_1^{(x)}=\frac{1}{Tp}\sum_{t=1}^T X_t^\top X_t, \qquad
\wh{\bSigma}_2^{(x)}=\frac{1}{Tq}\sum_{t=1}^T X_t X_t^\top, 
\end{equation}
\begin{equation} \label{est-1}
\wh\bV^{(1)}_{\tau,i,j}
= \{ \wh\bSigma_1^{(x)}\}^{-1/2} \sum_{t=1}^{T-\tau}
\frac{{ X}_{t+\tau}^\top \bE_{i,j}{ X_t} }{T-\tau}
\{ \wh\bSigma_1^{(x)}\}^{-1/2}, \;\;
\wh\bV^{(2)}_{\tau,i,j}
= \{ \wh\bSigma_2^{(x)}\}^{-1/2} \sum_{t=1}^{T-\tau} {{ X}_{t+\tau} \bE_{i,j}{ X_t}^\top \over T-\tau}
\{ \wh\bSigma_2^{(x)}\}^{-1/2},
\end{equation}
\begin{equation} \label{est-2}
\wh\bW^{(1)} = {1 \over p^2} \sum_{\tau=-\tau_0}^{\tau_0} \sum_{i=1}^p\sum_{j=1}^p
{\wh\bV^{(1)}_{\tau,i,j}\big(\wh\bV^{(1)}_{\tau,i,j}\big)^\top}, \qquad
\wh\bW^{(2)} = {1 \over q^2}\sum_{\tau=-\tau_0}^{\tau_0} \sum_{i=1}^q\sum_{j=1}^q
{\wh\bV^{(2)}_{\tau,i,j}\big(\wh\bV^{(2)}_{\tau,i,j}\big)^\top}.
\end{equation}
Performing the eigenanalysis for $\wh\bW^{(1)}$, and arranging the order of the resulting $q$ orthonormal eigenvectors $\wh \gamma_1, \cdots, \wh \gamma_q$ by the algorithm below, we take the
re-ordered eigenvectors as the columns of $\wh \bA_*$. The estimator $\wh \bB_*$ is obtained in the same manner via the eigenanalysis for $\wh\bW^{(2)}$. Then by (\ref{an1}), we obtain the transformed matrix series
\begin{equation} \label{est-3}
\wh \bU_t^* = \wh \bB_*^\top\{ \wh{\bSigma}_2^{(x)}\}^{-1/2}  X_t
\{\wh{\bSigma}_1^{(x)}\}^{-1/2} \wh \bA_*.
\end{equation}
Note that the estimation for $\bA_*$ and that for $\bB_*$ are carried out separately. They do not interfere with each other.

Now we present an algorithm to determine the order of the columns for $\wh \bA_*$. By (\ref{an1}), the columns of $ Z_t \equiv  X_t \{ \bSigma_1^{(x)}\}^{-1/2} \bA_*$ are divided into the $n_c$ uncorrelated blocks. Define
\begin{equation} \label{a12}
\wh \bZ_t \equiv  (\wh z_{t,i,j}) =  X_t \{ \wh \bSigma_1^{(x)}\}^{-1/2} (\wh \bgamma_1,
\cdots, \wh \bgamma_q).
\end{equation}
We divide the columns of $\wh \bZ_t $ into uncorrelated blocks according to the pairwise maximum cross correlations between the columns. Specifically, the maximum cross correlation between the $k$-th and $\ell$-th columns is defined as
\begin{align} \label{a13}
\wh \rho_{k,\ell}  = \max_{1\le i,j \le p, \, |\tau|\le \tau_1}
\big| \wh{\rm Corr} (\wh z_{t+\tau, i,k}, \, \wh z_{t,j, \ell} ) \big|
 = \max_{1\le i,j \le p, \, |\tau|\le \tau_1}
\frac{\big| \bgammahat_k^\top \bVhat^{(1)}_{\tau,i,j} \bgammahat_\ell\big|}
{\big\{\bgammahat_k^\top \bVhat^{(1)}_{0,i,i} \bgammahat_k \,
        \bgammahat_\ell^\top\bVhat^{(1)}_{0,j,j}
\bgammahat_\ell\big\}^{1/2}}
\end{align}
The second equality above follows from \eqref{est-1}, \eqref{est-2} and \eqref{a12}.
In the above expression, $\tau_1 \ge 1$ is a user-defined tuning parameter (See Remark \ref{remark:corPairs} below).
To determine all significantly correlated pairs of variable, rearrange $\wh\rho_{k,\ell}$, $1\le k < \ell \le q$, in the descending order: $\wh \rho_{(1)} \ge \cdots \ge \wh \rho_{(q_0)}$ and define
\begin{align}\label{a14mod}
\wh r=\arg\max_{1\le j\le q_0}  \frac{\wh \rho_{(j)}+\delta_T}{\wh \rho_{(j+1)}+\delta_T},
\end{align}
where $\delta_T>0$ is a small constant.
We take the $\wh r$ pair of columns corresponding to $\wh \rho_{(1)}, \cdots, \wh \rho_{(\wh r)}$ as correlated pairs, and treat the rest as uncorrelated pairs. The intuition is that $\rho_{(r)}/\rho_{(r+1)}$ is $\infty$ if $\rho_{(r)}> 0$ but $\rho_{(r+1)}=0$. The use of $\delta_T$ is to smooth out the large variation of $\wh \rho_{(j)}/\wh \rho_{(j+1)}$ when both $\rho_{(j)}$ and $\rho_{(j+1)}$ are small. Similar ideas have also been used in determining the number of factors in \cite{lam2012}, \cite{ahn2013} and \cite{han2022rank}.

With the $\wh r$ identified significantly correlated pairs, a connection graph is built, with the column indexes as the vertices and the edges between the identified correlated column pairs. The number of unconnected sub-graphs is the estimated number of blocks $\widehat{n}_c$, and each of maximum connected sub-graphs forms the estimated groups $\widehat{G}_i$, $i=1,\ldots \hat{n}_c$. The corresponding $\widehat{n}_c$ groups of $\wh \gamma_1, \cdots, \wh \gamma_q$ are taken as the columns of 
$\widehat A_{*i}$, $i=1,\ldots, \hat{n}_c$.
Algorithmically, 
%With the $\wh r$ identified correlated pairs, we group the columns of $\wh  Z_t$ as follows: 
%\begin{mydes} \item{(i)} S
start with $q$ groups with one column in each group; then iteratively check all pairs of groups and merge two groups together if there exists at least one pair of columns (one in each group) are significantly correlated; and stop when no groups can be merged.
%The number of unconnected groups is the estimated number of blocks $\widehat{n}_c$. 
%The corresponding $\widehat{n}_c$ groups of $\wh \gamma_1, \cdots, \wh \gamma_q$ are taken as the columns of $\widehat A_{*i}$, $i=1,\ldots, \hat{n}_c$.
%$\widehat{G}_i$, $i=1,\ldots, \hat{n}_c$. %, and the estimated transformation matrix for each group is $\widehat A_{*i}=(\widehat\gamma_k)_{k\in\widehat G_i}$.

The finite sample performance of the above algorithm  can be improved by prewhitening each column time series of $\wh \bZ_t$. This makes $\wh \rho_{k,\ell}$, for different $(k, \ell)$, more comparable. See Remark 2(iii) of \cite{chang2018}. In practice the prewhitening can be carried out by fitting each column time series a VAR model with the order between 0 and 5 determined by AIC. The resulting residual series is taken as a prewhitened series.

\begin{remark}
\label{remark:corPairs}
(i) The ratio estimator in (\ref{a14mod}) picks the $\wh r$ most correlated pairs of columns, and ignores the other small correlations in constructing the segmentation structure. Theorem \ref{th-3} in Section \ref{section:theories} below shows that the partition of $\wh A_*$ into $\{\wh A_{*1}, \cdots, \wh A_{*\wh n_c}\}$, determined by the above algorithm, provides a consistent estimation for the column segmentation of $\bA_*$.

%(ii) When $n_c=1$, i.e. all the $q$ columns of $U_t$ are correlated with each other, an approximate column segmentation will be obtained by ignoring the $(q_0-\wh r)$ smallest correlations.
%Empirical evidences indicate that such an approximation still leads to an improvement in future forecasting when $q$ is moderately large. See Section \ref{section:data} below and also the examples in \cite{chang2018}.

(ii) There are two tuning parameters used in the procedure, the maximum lag $\tau_0$ used in constructing $W^{(1)}$ and $W^{(2)}$ in \eqref{aa9} and the maximum lag $\tau_1$ used in measuring the dependency between two columns (rows) in \eqref{a13}. Lag $\tau_1$ is usually a sufficiently large integer, for example, between 10 and 20, in the spirit of the rule of thumb of \cite{box1970}.  Different values of $\tau_0$ and $\tau_1$ 
may, or may not, lead to different segmentation.
However the impact on, for example, future prediction is minimum, as the most information on linear dynamics is encapsulated in the most correlated pairs, as indicated by numerical examples in the Appendix.
\end{remark}

The ordering for columns of $\wh \bB_*$ is arranged, in the same manner as above, by examining the pairwise correlations among the columns of
\[
 X_t^\top \{ \wh \bSigma_2^{(x)}\}^{-1/2} (\wh \bgamma_1^{(2)},
\cdots, \wh \bgamma_p^{(2)}),
\]
where $\wh \bgamma_1^{(2)}, \cdots, \wh \bgamma_p^{(2)}$ are now the $p$ orthonormal eigenvectors of $\wh \bW^{(2)}$.

\section{Theoretical properties}\label{section:theories}

To gain more appreciation of the methodology, we will show the consistency of the proposed detection method of uncorrelated components. We mainly focus on the estimation error and the ordering of $\widehat A_*$, as those for $\widehat B_*$ are similar.
% Define 
% \begin{align*}
% \Sigma^{(1)}_{\tau,i,j}:=\E X_{t+\tau}^\top\bE_{ij} X_t,\qquad \bSigmahat^{(1)}_{\tau,i,j}:=\frac{1}{T-\tau}\sum_{t=1}^{T-\tau} X_{t+\tau}^\top\bE_{ij} X_t.    
% \end{align*}
Denote by $ X_{t,i,\cdot}$ the $i$-th row of $ X_t$, and $X_{t,i,j}$ the $(i,j)$-th element of $ X_t$.
For matrix $\bV=(V_{ij})$, let $\|\bV\|_{\rm op}$ denote the spectrum norm,
and $\|\bV\|_{\max} = \max_{i,j} |V_{ij}|$. We introduce some regularity conditions first.

\begin{assumption} \label{asmp:mixing}
Assume exponentially decay coefficients of the strong $\alpha$-mixing condition,
%\bes
$\alpha(k) \le \exp\big( - c_0 k^{r_1} \big)$
%\ees
for some constant $c_0>0$ and $0<r_1\le 1$, where
\begin{equation}\label{def:mixing}
\alpha(k) = \sup_t\Big\{\Big|\P(\cE_1\cap \cE_2) - \P(\cE_1)\P(\cE_2)\Big|:
\cE_1\in \sigma( X_s, s\le t), \cE_2\in \sigma( X_s, s\ge t+k)\Big\}.
\end{equation}
\end{assumption}

Here for any random variable/vector/matrix $ X$, $\sigma( X)$ is understood to be the $\sigma$-field generated by $ X$. Assumption \ref{asmp:mixing} allows a very general class of time series models, including causal ARMA processes with continuously distributed innovations; see \cite{bradley2005}, and also Section 2.6 of \cite{fan2008nonlinear}. The restriction $r_1\le 1$ is introduced only for simple presentation.

\begin{assumption} \label{asmp:eigenvalue}
Assume $\E X_t=0$. There exist certain finite constants $c_*, c^*$, such that
\begin{equation}
\sup_{i\le p, \|\bu\|_2=1} \Var\Big( X_{t,i,\cdot} \bu\Big) \le c^*,\quad
\inf_{\|\bu\|_2=1}\E \| X_t\bu \|_2^2/p\ \ge\ c_*.
\end{equation}
\end{assumption}
Note that $\sup_{ \|\bu\|_2=1}\Var\Big( X_{t,i,\cdot}^\top \bu\Big)= \|\E X_{t,i\cdot}^\top X_{t,i\cdot} \|_{\rm op}$ and $\E \| X_t\bu \|_2^2/p=u^\top \bSigma_1^{(x)}u$. Assumption \ref{asmp:eigenvalue} on the eigenvalues is a common assumption in the high dimensional setting, for instance, \cite{bickel2008covariance,bickel2008regularized, xia2017}.
% \Sigma^{(1)}_{0,i,i}
%It rules out the possibility of latent factor structure in $ X_t$. As the ultimate goal is forecasting, for ultra high dimensional matrix time series, we suggest to de-correlate the estimated latent processes after doing factor analysis first.

\begin{assumption}\label{asmp:tail}
For any $x>0$, 
%\begin{equation*}
$\max_{1\le i\le p,1\le j\le q} \P \left(|X_{t,i,j}|\ge x \right) \le c_1 \exp\left(-c_2 x^{r_2} \right) ,$  
%\end{equation*}
for some constant $c_1,c_2>0$ and $0<r_2\le 2$.
\end{assumption}

Assumption \ref{asmp:tail} requires that the tail probability of each individual series of $ X_t$ decay exponentially fast. In particular, when $r_2=2$, each $X_{t,i,j}$ is sub-Gaussian.

Write the eigenvalue-eigenvector decomposition of $\bW^{(1)}$ as
\bes
\bW^{(1)} = \bGamma^{(1)}\bD\big(\bGamma^{(1)}\big)^\top,
\ees
where $\bD=\diag(\lam_1,\ldots,\lam_q)$ %contains the eigenvalues
with $\lam_1\ge \cdots \ge \lam_q$, and $\bGamma^{(1)} = (\bgamma_1,\ldots,\bgamma_q)$. By the structure assumption in (\ref{a1}),
the index set $\{1, \cdots, q\}$ is partitioned
into $n_c$ subsets
$ G_1,...,G_{n_c}$ such that {the columns of different sub-matrices}
%\bel{gamma-decomp}
$ X_t\Big(\bSigma^{(1)}_1\Big)^{-1/2}\bGamma^{(1)}_{G_i},\ i = 1,\ldots,n_c,$
%\eel
are uncorrelated with each other across all time lags. 
Define eigen-gap
\begin{align}\label{eigen-gap}
\Delta=\min_{1\le i<j\le n_c} \min_{k\in G_i, \ell\in G_j} \left| \lambda_{k}-\lambda_{\ell}\right|.   
\end{align}
%where $(G_1,...,G_{n_c})$ are specified above.}
%\yfm{Let the estimated transformation matrix for each oracle group be $\widehat A_{*j}=(\widehat\gamma_k)_{k\in G_j}$.}

The following theorem provides the non-asymptotic bounds for the estimators of $\bV^{(1)}_{\tau,i,j}$, $\bW^{(1)}$ {and $ A_{*j}$}, $1\le i,j\le q$ under both exponential decay $\alpha$ mixing condition and exponential tail condition of $ X_t$. 

\begin{theorem}\label{th-1new}
Suppose Assumptions \ref{asmp:mixing}, \ref{asmp:eigenvalue}, \ref{asmp:tail} hold with constants $c_*,c^*, r_1,r_2$, and $\tau_0$ is a finite constant. Let $1/\beta_1=1/r_1+2/r_2$ and $1/\beta_2=1/r_1+1/r_2$. Then,
\begin{align}\label{th-1-1new}
% &\Big\|\bSigmahat^{(1)}_{\tau,i,j} - \bSigma^{(1)}_{\tau,i,j}\Big\|_{\rm op}
% %\le q\Big\|\bSigmahat^{(1)}_{\tau,i,j} - \bSigma^{(1)}_{\tau,i,j}\Big\|_{\max}
% \le C_1 \eta_{T,p,q}, \;\; 
&\big\|\bVhat^{(1)}_{\tau,i,j} - \bV^{(1)}_{\tau,i,j}\big\|_{\rm op} \le C_1 \eta_{T,p,q}, \;\; \forall 1\le \tau\le \tau_0, 1\le i,j\le p, \\
& \big\|\bWhat^{(1)}  - \bW^{(1)} \big\|_{\rm op} \le C_1 \eta_{T,p,q}, \label{th-1b-1new}
%\ \forall 0\le \tau\le\tau_0,i\le p,j\le q
%\Big\}
%\ge 1-2\eps_T
\end{align}
in an event $\Omega_T$ with probability at least $1 -\eps_T$, where
\bel{th-1-2new}
\eta_{T,p,q} = q\left(\sqrt{\frac{\log(pq/\epsilon_T)}{T}} + \frac{[\log(Tpq/\epsilon_T)]^{1/\beta_1}} {T} + \frac{[\log(Tpq/\epsilon_T)]^{2/\beta_2}} {T^2}\right),
\eel
%$C_1$ is a positive constant depending on $r_1,r_2$, and 
$C_1$ is a constant depending on $c_*,c^*,r_1,r_2$ only. Moreover, there exists $\wt A_* \equiv (\wt A_{*1}, \cdots,
\wt A_{*n_c}) $ of which the columns are a permutation
of $(\wh \gamma_1, \cdots, \wh\gamma_q)$ such that
%if $C_1 \eta_{T,p,q} \le \Delta/2$, then
\bel{th-1-3new}
% && \big\|\bVhat^{(1)}_{\tau,i,j} - \bV^{(1)}_{\tau,i,j}\big\|_{\rm op} \le C_2 \eta_{T,p,q}, \qquad 
% \big\|\bWhat^{(1)}  - \bW^{(1)} \big\|_{\rm op} \le C_2 \eta_{T,p,q},
\big\|\wt A_{*j} \wt A_{*j}^\top  - A_{*j} A_{*j}^\top \big\|_{\rm op} \le C_2 \eta_{T,p,q} \quad {\rm for}
\quad 1\le j\le n_c
\eel
in the same event $\Omega_T$ provided $C_1 \eta_{T,p,q} \le \Delta/2$, where $C_2$ is a constant depending on $c_*,c^*,r_1,r_2$ only. 
\end{theorem}

\begin{remark} \label{rm-rates}
%Suppose the eigenvalues of $W^{(1)}$ and $W^{(2)}$ are all distinct. 
(i) Let $P_G$ be the projection operator onto the column space of $G$, which can be written as $P_G=G (G^\top G)^{-1} G^\top$.
Under the conditions of Theorem \ref{th-1new}, we can obtain {$\| P_{\wt A_{*j}\otimes \wt B_{*i}} - P_{A_{*j}\otimes B_{*i}}\|_{\rm op}=O_{\P} (\max\{p,q\} [(\log(pq)/T)^{1/2} + \log(Tpq)^{1/\beta_1}/T])$ for each sub-group $1\le i\le n_r, 1\le j\le n_c$}, where the columns of $\wt B_*$ are a permutation of the
orthonormal eigenvectors of ${\wh W}_{(2)}$. If we stack all the elements of $X_t$ into a long vector such that $\vec( X_t) = \bPhi \vec(\bU_t)$ and apply TS-PCA of \cite{chang2018} directly, the estimation of the decorrelation transformation matrix would satisfy {$\| P_{\wt\Phi_k} - P_{\Phi_k}\|_{\rm op}=O_{\P} (pq[(\log(pq)/T)^{1/2} + \log(Tpq)^{1/\beta_1}/T])$ for $1\le k\le n_r n_c$ and $\Phi=(\Phi_1,...,\Phi_{n_r n_c})$}. Obviously, $\| P_{\wt A_{*j}\otimes \wt B_{*i}} - P_{A_{*j}\otimes B_{*i}}\|_{\rm op}$ has much sharper rate which is equivalent to that for the TS-PCA estimation with dimension max$(p,q)$. 

(ii) The consistency of $\wh W^{(1)}$ requires $\max(p,q)=o(\sqrt{T})$. When $A$ and $B$, or equivalently $W^{(1)}$ and $W^{(2)}$, have certain sparsity structures, this condition can be further relaxed. In TS-PCA, threshold estimator of \cite{bickel2008covariance} is employed to construct a sparse $\wh W^{(1)}$, and then the convergence rates are improved to allow the dimension to be much larger than $T$ \citep{chang2018}. 
The similar extension can be established in our setting.
%In this paper, we do not pursue this direction, as both the threshold estimators and the \rc{corresponding} theoretical analysis can be extended from \cite{chang2018}.}
% \rc{In this case}, \yfm{%Inheriting the spirit of threshold estimator for large covariance matrix by \cite{bickel2008covariance}, 
% threshold estimator of \cite{bickel2008covariance} can be employed to construct \rc{a sparse} $\wh W^{(1)}$. The convergence rates can thus be improved to allow $\max(p,q)\gg T$. In this paper, we do not pursue this direction, as both the threshold estimators and the \rc{corresponding} theoretical analysis will be similar to those in \cite{chang2018}.}
\end{remark}

When the decays of the $\alpha$-mixing coefficients and the tail probabilities of $ X_t$ are slower than Assumptions \ref{asmp:mixing} and \ref{asmp:tail}, we impose the following Assumptions \ref{asmp:mixing2} and \ref{asmp:tail2} {instead}. These conditions ensure the Fuk-Nagaev-type inequalities for $\alpha$-mixing processes; see also \cite{rlo2000theorie}, \cite{liu2013probability, wu2016performance} and \cite{zhang2017gaussian}.

\begin{assumption} \label{asmp:mixing2}
%Assume algebraically decay coefficients of the strong
Let the $\alpha$-mixing coefficients satisfy the condition
%\bes
$\alpha(k) \le c_0 k^{-r_1},$
%\ees
where $\alpha(k)$ is defined in \eqref{def:mixing}, $c_0>0$ and $r_1>1$.
\end{assumption}

\begin{assumption}\label{asmp:tail2}
For any $x>0$, 
%\begin{equation*}
$\max_{1\le i\le p,1\le j\le q} \P \left(|X_{t,i,j}|\ge x \right) \le c_1 x^{-2r_2} ,$  
%\end{equation*}
for some constant $c_1>0$ and $r_2>1$.
\end{assumption}

Theorem \ref{th-1-poly} below presents the uniform convergence rate for $\bVhat^{(1)}_{\tau,i,j}$ and $\bWhat^{(1)}$, $1\le i,j\le q$. It is intuitively clear that the rate is much slower than that in Theorem \ref{th-1new}.

\begin{theorem}\label{th-1-poly}
Suppose Assumptions \ref{asmp:eigenvalue}, \ref{asmp:mixing2}, \ref{asmp:tail2} hold with constants $c_*,c^*, r_1,r_2$, and $\tau_0$ is a finite constant. Let $\beta_3=r_2(r_1+1)/(r_1+r_2)>1$. Then, \eqref{th-1-1new} and \eqref{th-1b-1new} hold in an event $\Omega_T$ with probability at least $1-\epsilon_T$, where
\bel{th-1-1poly}
\eta_{T,p,q} = q\left( \frac{(Tpq/\epsilon_T)^{1/\beta_3}}{T} + \sqrt{\frac{\log(pq/\epsilon_T)}{T}} \right) ,
\eel
and $C_1$ depends on $c_*,c^*,r_1,r_2$ only. Similarly, if $C_1 \eta_{T,p,q} \le \Delta/2$, then there exists $\wt A_* \equiv (\wt A_{*1}, \cdots,
\wt A_{*n_c}) $ of which the columns are a permutation of $(\wh \gamma_1, \cdots, \wh\gamma_q)$ such that \eqref{th-1-3new} holds in the same event $\Omega_T$ for some positive constant $C_2$ depending on $c_*,c^*,r_1,r_2$ only.
\end{theorem}

Theorem \ref{th-3} below implies that the partition
of $\wh A_*$ into $\{ \wh A_{1*}, \cdots, \wh A_{*\wh n_c}\}$ in Section \ref{section:estimation}
provides a consistent estimation
for the column segmentation of $A_*
=(A_{*1}, \cdots, A_{*n_c})$. To this end, let
\bel{inseparable0}
\rho_{k,\ell}= \max_{1\le i,j\le p,|\tau|\le\tau_1}
\frac{\big| \bgamma_k^\top \bV^{(1)}_{\tau,i,j} \bgamma_\ell\big|}
{\big\{\bgamma_k^\top \bV^{(1)}_{0,i,i} \bgamma_k\,
         \bgamma_\ell^\top \bV^{(1)}_{0,j,j} \bgamma_\ell \big\}^{1/2}}.
\eel
See also (\ref{a13}).
By (\ref{an1}), the columns of $ Z_t \equiv  X_t \{ \bSigma_1^{(x)}\}^{-1/2} \bA_*$ are divided into the $n_c$ uncorrelated blocks. 
We assume that those $n_c$ blocks can also be obtained by
the algorithm in Section \ref{section:estimation}
with $\wh \rho_{k, \ell}$ replaced by $\rho_{k,\ell}\cdot I(\rho_{k,\ell}\ge \rho)$ for some constant $\rho>0$.
Denote by $G_1, \cdots, G_{n_c}$, respectively, the indices of the components of $Z_t$ in each of those
$n_c$ blocks. Denote by $\wh G_1, \cdots, \wh G_{\wh n_c}$, respectively, the indices of the components
of $\wh Z_t$ in each of the $\wh n_c$ uncorrelated blocks identified by the algorithm in Section \ref{section:estimation}. Re-arrange of the order of
$\wh G_1, \cdots, \wh G_{\wh n_c}$ if necessary. Then the theorem below implies that
$\P(\hat n_c =n_c) \to 1$ and $\P(\wh G_i = G_i | \hat n_c = n_c) \to 1$ for $1\le i \le n_c$.

%The following theorem states the consistency of the detection method in Section \ref{section:estimation}.
% We say that a group $G_{i_0}$ in (\ref{gamma-decomp}) is {\it detected} if
% \bel{detect}
% \Ghat_j=G_{i_0}, 
% \eel
% for some $\ 1\le j\le\wh n_c$. 
% The following theorem spells out the regularity conditions required for (\ref{detect})
% in our analysis.

\begin{theorem}\label{th-3}
Suppose conditions of Theorem \ref{th-1new} or \ref{th-1-poly} hold and $\tau_1$ is a finite constant. 
%is a true partition of $\{1,...,q\}$. Assume every group $G_{i}$ is inseparable at level $\rho^*>0$. Let $(\widehat{G}_1,...,\widehat{G}_{\widehat{n}_c})$ be the estimated group using $\wh r$ defined in \eqref{a14mod} with $\delta_T\le (\rho^{*})^2/2$.
Suppose
\begin{align}\label{cond}
\kappa_1 \eta_{T,p,q}<\Delta, \qquad     \kappa_2 \eta_{T,p,q} < \rho^* ,
\end{align}
where $\kappa_1,\kappa_2$ are certain positive constants depending on $c_*,c^*,r_1,r_2$ only, $\eta_{T,p,q}$ is defined in \eqref{th-1-2new} or \eqref{th-1-1poly} and depends on $\epsilon_T$. Then in an event $\Omega_T$ with probability at least $1 -\eps_T$, we have 
\[
\wh n_c=n_c \mbox{\ \ and \ \ }
%the estimated groups satisfy 
\wh G_i=G_i,\quad 1\le i\le n_c.
\]
%up to a permutation of group indices and within group column indices. 
\end{theorem}

\begin{remark}
The first inequality in \eqref{cond} requires that the minimum eigen-gap $\Delta$ between different uncorrelated groups is sufficiently larger than the estimation error $\eta_{T,p,q}$, such that all the groups are identifiable. The second inequality in \eqref{cond} ensures that there are no cross-group edges among  $G_1, \cdots, G_{n_c}$.
%$i=1,\ldots, n_c$, when using the maximum cross correlation detection method. 
The constants $\kappa_1$ and $\kappa_2$ are specified in the proof of the theorem.
\end{remark}

\section{Numerical properties}\label{sec4}

\subsection{Simulation} \label{section:simulation}

We illustrate the proposed decorrelation method with simulated examples. We always set $\tau_0=5$ in (\ref{est-2}), and $\tau_1 =15$ in (\ref{a13}).
The prewhitening, as stated at the end of Section \ref{section:estimation}, is always applied in determining the groups of the columns of $\wh \bA_*$ and $\wh \bB_*$. As the true transformation matrices $\bA_*$ and $\bB_*$ defined in (\ref{an2}) cannot be computed easily even for simulated examples, we use the following proxies instead 
\[
\bA_*= \{ \wh \bSigma_1^{(x)}\}^{-1/2} \bA \{ \wh \bSigma_1^{(u)} \}^{1/2}, \qquad
\bB_*= \{ \wh \bSigma_2^{(x)}\}^{-1/2} \bB \{ \wh \bSigma_2^{(u)} \}^{1/2}, 
\]
where $\wh \bSigma_1^{(x)}, \; \wh \bSigma_2^{(x)}$ are given in (\ref{est-0}), and
\[
\wh \bSigma_1^{(u)}= {1 \over T p} \sum_{t=1}^T \bU_t^\top \bB^\top \bB \bU_t, \qquad
\wh \bSigma_2^{(u)}= {1 \over T q} \sum_{t=1}^T \bU_t \bA \bA^\top \bU_t^\top. 
\]
%({\color{red} Is this correct? In Example 3, you seem to use the true $\bSigma_1^{(x)}$ etc instead in some oracle models. I can guess why. But looks inconsistent. An easy way out: delete (O2) from Example 3, which adds little insight anyway -- QY.})
To abuse the notation, we still use $\bA_*, \; \bB_*$ to denote their proxies in this section since the true $\bA_*, \; \bB_*$ never occur in our computation. For each setting, we replicate the simulation 1000 times.

%In this section, we conduct simulation studies to illustrate the finite sample performance of the proposed method using three examples. To identifying the correlated columns of $\topc  X_t$ (see \eqref{a12}) or the correlated rows of $\topr  X_t$ (similar to \eqref{a12} for row segmentation), we pre-whiten each transformed columns of $\topc  X_t$ or rows of $\topr  X_t$ before applying the detection methods described in Section \ref{section:estimation}. The prewhitening is carried out by fitting each column series 
%of $\topc  X_t$ (or row series of $\topr  X_t$) a VAR model with the order determined by AIC but not greater than 5. We then use the resulting residual series to calculate the maximum correlations with proper $\tau_1$ in (\ref{a13}). 

\begin{example}
We consider model (\ref{a1}) in which all the component series of $\bU_t$ are independent and ARMA(1, 2) of the form:
\begin{equation} \label{e1}
z_t = b z_{t-1} + \epsilon_t + a_1 \epsilon_{t-1} + a_2 \epsilon_{t-2},
\end{equation}
where $b$ is drawn independently from the uniform distribution on $(-.98,\,-0.5)\cup (0.5,\, 0.98)$, $a_1, a_2$ are drawn independently from the uniform distribution on $(-.98,\,-0.3)\cup (0.3,\, 0.98)$, and $\epsilon_t$ are independent and $N(0,1)$. The elements of $\bA$ and $\bB$ are drawn independently from $U(-1, \, 1)$. For this example, we assume that we know %$p_i=q_i=1$,
$n_r=p$ and $n_c=q$ in (\ref{a1}). The focus is to investigate the impact of $p$ and $q$ on the estimation for $\bB$ and $\bA$.

% Setting $\tau_0=5$, we perform the eigen-decomposition for matrix $\bWhat^{(x)}$ and obtain its $q$ orthonormal eigenvectors $\bGammahat^{(x)}$. In practice we need to 
% estimate the permutation matrix $\bUpsilon$ and then measure the error between 
% $\wh \bA_*=\bGammahat^{(x)}\hat{\bUpsilon}$ and $\widetilde\bA_{\small\rm true}$. To simplify the evaluation, we notice that $\wt\bA_{\small \rm true}^\top\bGamma^{(x)}=\bUpsilon$ is a permutation matrix,  
% in which each row or column is a unit vector with
% only one element equal to 1. Therefore we can measure the estimation error directly
% (i.e. without the required estimation of column permutation) by measuring 
% the distance between $\wt\bA_{\small \rm true}^\top\bGammahat^{(x)}$ and its closest 
% permutation matrix as a proxy for the distance between $\wh \bA_*$ and $\wt\bA_{\small\rm true}$, by
Performing the eigenanalysis for $\wh \bW^{(1)}$ and $\wh \bW^{(2)}$ defined in (\ref{est-2}), we take the resulting two sets of orthonormal eigenvectors as the columns of $\wh \bA_*$ and $\wh \bB_*$ respectively. To measure the estimation error, we define
\begin{align*}
D(\wh \bA_*, \, \bA_*) = \frac{1}{2 q (\sqrt{q} -1)} \sum_{j=1}^q \big( \frac{1}{\max_i |d_{i,j}|}
+  \frac{1}{\max_i |d_{j,i}|}
-2 \big),
\end{align*}
where $d_{i,j}$ is the $(i,j)$-th element of matrix $\wh\bA_{*}^{\top} \bA_{*}$. Note that $D(\wh \bA_*, \, \bA_{*})$ is always between 0 and 1,
and $D(\wh\bA_*, \, \bA_{*})=0$ if $\wh\bA_*$ is a column permutation of $\bA_{*}$. % $D(\wh \bB_*, \, \bB_*)$ is defined in the same manner.

%\rcqq{For Qiwei: The criterion above does not prevent you to have two 1's in a column, or one (i,j) used twice. Can we have an operational definition? One simple way is "Find the maximum. Remove it's corresponding row and column, repeat.". Fancy way: make it an optimization problem. "All ones on diagonal, switching row and column repeatedly." }{\color{red} The orthogonality of the two matrices ensures no columns have two 1 -- QY}

\begin{table}[ht]\caption{Example 1 -- Means and standard errors (SE) of $D(\wh\bA_*, \, \bA_{*})$ and $D(\wh\bB_*, \, \bB_{*})$ in a simulation with 1000 replications.} \label{tab1}
%\small \footnotesize
\scriptsize
\begin{center}
\begin{tabular}{r|c|cc|cc||c|cc|cc|}
 &  & \multicolumn{2}{c|}{$D(\wh\bA_*, \,
\bA_{*})$} & 
\multicolumn{2}{c||}{$D(\wh\bB_*, \, \bB_{*})$}
&  & \multicolumn{2}{c|}{$D(\wh\bA_*, \,
\bA_{*})$} &
\multicolumn{2}{c|}{$D(\wh\bB_*, \, \bB_{*})$}\\
$T$ & ($q,\; p$) & Mean& SE & Mean & SE & ($q,\; p$) & Mean& SE & Mean & SE\\
\hline
100 & (4, 4) & 0.135 & 0.103 & 0.116 & 0.094 & (4, 8) & 0.124 & 0.096& 0.169 & 0.056 \\
500 & & 0.081 & 0.080 & 0.052 & 0.065 &  &  0.074 & 0.078& 0.089& 0.047 \\
1000& & 0.048 & 0.062 & 0.050 & 0.063 &  & 0.057 & 0.066& 0.056& 0.037 \\
5000& & 0.021 & 0.042 & 0.020 & 0.042 & & 0.019 & 0.039& 0.032& 0.027 \\

\hline
100 & (8, 8)& 0.172 & 0.057 & 0.172 & 0.055 & (16, 16) & 0.204 & 0.031 & 0.206 & 0.032 \\
500 & & 0.098 & 0.048 & 0.086 & 0.044 & & 0.132 & 0.032 & 0.131 & 0.031 \\
1000& & 0.074 & 0.045 & 0.070 & 0.041 & & 0.097 & 0.027 & 0.093 & 0.026 \\
5000& & 0.034 & 0.029 & 0.035 & 0.030 & & 0.041 & 0.017 & 0.042 & 0.018 \\
\hline
100& (32, 32) & 0.224 & 0.018 & 0.225 & 0.018 & (100, 16)&  0.251 & 0.006 & 0.187 & 0.032 \\
500 & & 0.162 & 0.019 & 0.161 & 0.020 & & 0.212 & 0.008 & 0.127 & 0.031 \\
1000& & 0.127 & 0.018 & 0.125 & 0.018 & & 0.185 & 0.009 & 0.099 & 0.028 \\
5000& & 0.058 & 0.013 & 0.058 & 0.013 & & 0.106 & 0.008 & 0.044 & 0.019 \\
\end{tabular}
\end{center}
\end{table}

We set $T=100, 500, 1000, 5000$, and $p, q = 4, 8, 16, 32, 100$. The means and standard errors of $D(\wh\bA_*, \, \bA_{*})$ and $D(\wh\bB_*, \, \bB_{*})$ over the 1000 replications are reported in Table \ref{tab1}. As expected, the estimation errors decrease as $T$ increases. Furthermore the error in estimating $\bA_*$ increases as $q$ increases, and that in estimating $\bB_*$ increases as $p$ increases. Also noticeable is the fact that the quality of the estimation for $\bA_*$ depends on $q$ only and that for $\bB_*$ depends on $p$ only, as $D(\wh\bA_*, \, \bA_{*})$ is about the same for $(q,p)=(4, 4)$ and $(q, p)=(4,8)$, and $D(\wh\bB_*, \, \bB_{*})$ is about the same for $(q, p)=(4,8)$ and $(q, p)=(8,8)$. When $(q,p)=(32, 32)$, each $\bA$ and $\bB$ contains $1024$ unknown parameters. The quality of their estimation with $T=100$ is not ideal, but not substantially worse than that for $(q,p)=(16, 16)$, and the estimation is very accurate with $T=5000$. For $(q,p)=(100, 16)$, matrix time series $ X_t$ contains 1600 component series. The estimation for $\bA_*$ is not accurate enough as $q=100$, while the estimation for $\bB_*$ remains as good as that for $(q,p)=(16,16)$, and is clearly better than that for $(q,p)=(32,32)$.
\end{example}

% \bigskip

\begin{example}
%\noindent {\bf Example 2}. 
To examine the performance of the proposed method for identifying uncorrelated blocks, we consider now a only column-segmentation model
\[
 X_t =  \bU_t \bA^\top,
\]
where $ X_t, \bU_t$ are $p\times q$ matrix time series. Note that adding the row transformation matrix $\bB$ in the model entails the application of the same method twice without adding any new insights on the performance of the method, as the estimation for $\bA_*$ and that for $\bB_*$ are performed separately. See Section \ref{section:estimation}. 

The transformation matrix $\bA$ in the above model is generated in the same way as in Example 1 above. All the element time series %Columns $1, 4, 6, 7\cdots, q$ 
of $\bU_t$, except those in columns 2, 3 and 5, are simulated independently from ARMA(1,2) model (\ref{e1}). Denoted by $\bU_{t, j}$ the $j$-th column of $\bU_t$,
we let
\[
\bU_{t,2}=\bU_{t+1,1}, \quad \bU_{t,3}=\bU_{t+2,1}, \quad 
 \bU_{t,5}=\bU_{t+1,4}.
%\[
%\bZ_{t, j+1} = \bZ_{t+j, 1} \;\; {\rm for} \; j=1, 2 \quad {\rm and} \quad
%\bZ_{t, 5} = \bZ_{t+1, 4},
\]
% where $\bU_{t, j}$ denotes the $j$-th column of $\bU_t$.
Hence the first three columns of $\bU_t$ forms one block of size 3, columns 4 and 5 forms a block of size 2, and the rest of the columns are
uncorrelated with each other, and are uncorrelated to the first two blocks.

%With $\tau_0=5$ we compute $\wh \bW^{(x)}$ and its corresponding $\bGammahat^{(x)}$ according to (\ref{est-2}) and \eqref{est-22}. We choose to use $\tau_1 =15$ in this example. Applying ratio-estimate (\ref{a14}) to determine all correlated column pairs, we form the grouping according to the three steps in Section \ref{section:estimation}.

We set $T=100, 500, 1000, 5000$. For each sample size, we set $(p,q)=(3,6)$, $(6,6)$, $(6, 10)$ and $(10, 10)$. Hence the number of groups is $n_c=3,3,7$ and $7$, respectively. For each setting, we perform the eigenanalysis for $\wh \bW^{(1)}$ defined as (\ref{est-2}). We then apply the algorithm in Section \ref{section:estimation} to arrange the orders of the $q$ resulting orthonormal eigenvectors to form estimator $\wh \bA_*$, for which the number of the correlated pairs of columns is determined by (\ref{a14mod}).
% For each setting, we replicate the estimation 1000 times.
Table \ref{tab2} reports the relative frequencies of the correct specification of all the $n_c$ uncorrelated blocks. We also report the relative frequencies of two types of wrong segmentation: (a) Merging with $\wh n_c = n_c -1$, i.e. $n_c-2$ blocks are correctly specified, and the remaining two blocks are put into one; (b) Splitting with $\wh n_c = n_c+1$, i.e. $n_c -1$ blocks are correctly specified, and the remaining block
incorrectly splits into two. Table \ref{tab2} indicates that the relative frequency of the correct specification increases as the sample size $T$ increases, and it decreases as $q$ increases while the performance is much less sensitive to the increase of $p$.
% When $\wh n_c = n_c -1$, two uncorrelated groups are mistakenly put together.
% When $\wh n_c = n_c +1$, one of the two groups with more than one columns each
% is wrongly splitted into two groups.
It is noticeable that the probability of the event $\{ \wh n_c = n_c +1 \}$ is very small especially when $T$ is large. With $q=6$, the probability for the correct specification is not smaller than 64\% with $T=500$, and is greater than 70\% with $T=1000$. Furthermore the probability for $\wh n_c = n_c$ or $n_c-1$ is over 92\% for $T\ge 500$. Note that when $\wh n_c = n_c-1$, an effective dimension reduction is still achieved in spite of missing a split.

For the instances with the correct specification, we also calculate the estimation errors for the $n_c$ subspaces. More precisely, put $\bA_{*} = (\bA_{1}, \cdots, \bA_{n_c}) $ and $\wh \bA_* = (\wh \bA_{1}, \cdots, \wh \bA_{n_c})$. We measure the estimation error by
\begin{align*}
D_1 (\wh \bA_*, \bA_{*}) = \frac{1}{n_c} \sum_{j=1}^{n_c}
\big\{ 1 - \frac{1}{{\rm rank}(\bA_{j})} {\rm trace}\big(\bA_{j} \bA_{j}^{\top} \wh\bA_{j} \wh\bA_{j}^{\top}\big) \big\}.
\end{align*}
Note that $\{{\rm rank}(\bA_{j})\}^{-1} {\rm trace}\big(\bA_{j} \bA_{j}^{\top} \wh\bA_{j} \wh\bA_{j}^{\top}\big)$ is always between 0 and 1. Furthermore it is equal to 1 if $\calM(\bA_{j}) = \calM(\wh \bA_{j})$, and 0 if the two spaces are perpendicular with each other; see, for example, \cite{stewart1990} and \cite{pan2008}. The boxplots of $D_1 (\wh\bA_{*}, \bA_{*}) $ presented in Figure \ref{fig1} indicate that the estimation errors decrease when $T$ increases, and the errors with large $q$ are greater than those with small $q$.

\begin{table}[ht]
\caption{Example 2 -- Relative frequencies correct and incorrect segmentation in a simulation with 1000 replications. 
%for both the correct specification (CS) (i.e. $\wh n_c = n_c$) and the incomplete specification but with $\wh n_c = n_c -1$  of the uncorrelated columns groups in a simulation with 1000 replications. For this example $n_c = q-3$.
} \label{tab2}
%\small
\scriptsize
\begin{center}
\begin{tabular}{c|c|cccc||c|cccccccccc}
$T$& $(q,p)$ %\multicolumn{4}{c}{$(6,3)$} & \multicolumn{4}{c}{$(6,6)$} & \multicolumn{4}{c}{$(10,6)$} & \multicolumn{4}{c}{$(10,10)$} \\ \hline\hline
 & Correct & Merging & Splitting & Other & $(q,p)$ & Correct & Merging & Splitting & Other  \\\hline
100 & $(6,3)$ & 0.369 & 0.274 & 0.299 & 0.058 & $(6,6)$ & 0.371 & 0.207 & 0.208 & 0.214 \\
500 &         & 0.659 & 0.237 & 0.056 & 0.048 &         & 0.639 & 0.218 & 0.071 & 0.072 \\
1000 &        & 0.721 & 0.220 & 0.031 & 0.028 &         & 0.706 & 0.219 & 0.030 & 0.045 \\
5000 &        & 0.837 & 0.142 & 0.008 & 0.013 &         & 0.815 & 0.156 & 0.011 & 0.018 \\ \hline
100 & $(10,6)$& 0.103 & 0.087 & 0.270 & 0.540 &$(10,10)$& 0.116 & 0.091 & 0.249 & 0.544 \\
500 &         & 0.323 & 0.148 & 0.141 & 0.388 &         & 0.299 & 0.195 & 0.116 & 0.390 \\
1000 &        & 0.369 & 0.227 & 0.086 & 0.318 &         & 0.363 & 0.218 & 0.067 & 0.352 \\
5000 &        & 0.542 & 0.248 & 0.014 & 0.196 &         & 0.495 & 0.269 & 0.019 & 0.217 \\
\end{tabular}
\end{center}
\end{table}

\begin{figure}[ht]
%\begin{center}\includegraphics[scale=0.7]{fig1}\end{center}
\begin{center}\includegraphics[scale=0.4]{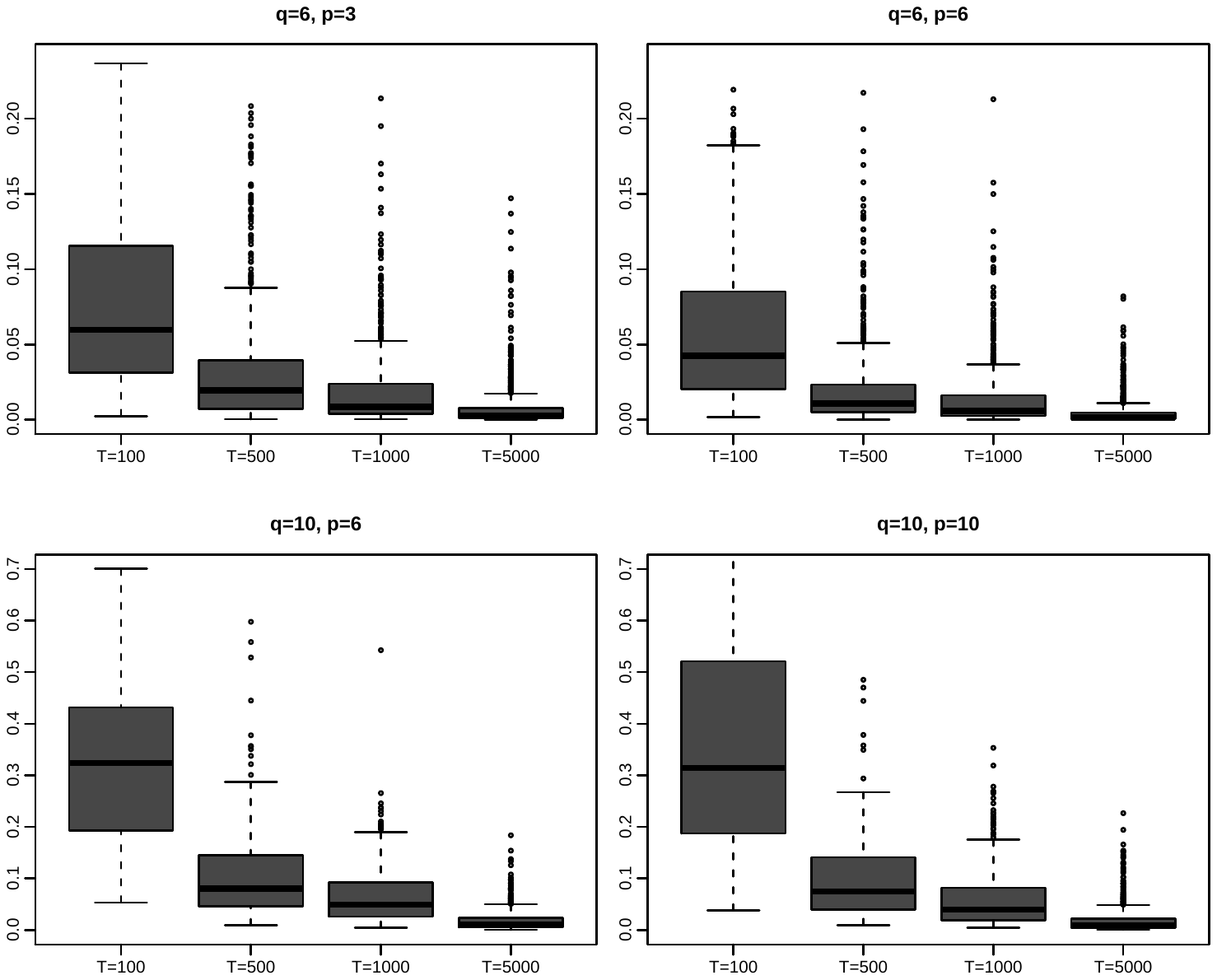}\end{center}
\caption{Example 2 -- boxplots of $D_1(\wh \bA_*, \bA_{*})$ in a simulation
with 1000 replications. %\rcqq{Smaller figure, one row. Large font on the title (or move them inside the boxes.}
} \label{fig1}
\end{figure}

\end{example}

%\noindent {\bf Example 3}. 
\begin{example}
In this example we examine the gain in post-sample forecasting due to the decorrelation transformation. Now in model (\ref{a1}) all the elements of $\bA$ and $\bB$ are drawn independently from $U(-1, \, 1)$, with $(p,q)$ equal to $(6,6)$, $(6,8)$, $(8,8)$ and $(10,10)$. For each $(p,q)$ combination, the first three rows (or columns) form one block, the next two rows (or columns) form a second block, and all the other 
row (or column) is independent of the rest of the rows (or columns). Table~\ref{table:example3} shows the configuration of $\bU_t$ for $(p,q)=(8,8)$.

\begin{table}\label{table:example3}
%\small
%\footnotesize
%\scriptsize{
\tiny{
\begin{center}
\begin{tabular}{c|ccc|cc|c|c|c|}
 & 1 & 2 & 3 & 4 & 5 & 6 & 7 & 8 \\ \hline
 1 & \multicolumn{3}{|c|}{\multirow{3}{*}{.81}} &  \multicolumn{2}{|c|}{\multirow{3}{*}{.64}} & & & \\
 2 & \multicolumn{3}{|c|}{} & \multicolumn{2}{|c|}{} & & & \\
 3 & \multicolumn{3}{|c|}{} & \multicolumn{2}{|c|}{} & & & \\ \hline
 %2 & \multicolumn{3}{c|}{.81} & \multicolumn{2}{c|}{.64}& & &  \\
 %4 & \multicolumn{3}{c|}{.25}& \multicolumn{2}{c|}{.25} & & &  \\
 4 & \multicolumn{3}{|c|}{\multirow{2}{*}{.25}} &  \multicolumn{2}{|c|}{\multirow{2}{*}{.25}} & & & \\
 5 & \multicolumn{3}{|c|}{} & \multicolumn{2}{|c|}{} & & & \\ \hline
 6 & & & & & & & &  \\ \hline
 7 & & & & & & & &  \\ \hline
 8 & & & & & & & &  \\ \hline
\end{tabular}
\end{center}
\caption{Example 3: Block configuration of $\bU_t$ for the case of $(p,q)=(8,8)$} 
}
\end{table}

Each univariate block of $\bU_t$ (e.g. $U_{t,6,6}$) is an AR(1) process with the autoregressive coefficient drawn from the $U[0.1,0.3]$ and
independent and $N(0,1)$ innovations. For each block of size $(3,1)$, $(1,3)$, $(2,1)$ and $(1,2)$, a vector AR(1) model is used. All elements
of the coefficient matrix is drawn independently from $U[0,1]$, then normalized so its singular values are within $[0.1, 0.3]$. The vector innovations consist of independent and N$(0,1)$ random variables. Each of the remaining four blocks follow a matrix AR(1) (i.e. MAR(1))  model of \cite{chen2021autoregressive}, i.e. 
\begin{align}\label{model:ex3}
\bU_{t,i,j}=\lambda_{i,j}\bPhi_{1,i,j} \bU_{t-1,i,j} \bPhi_{2,i,j}^\top+\bE_{t,i,j}, \quad
i,j =1, 2,
\end{align}
where $\bPhi_{1,i,j}$ and $\bPhi_{2,i,j}$ are the coefficient matrices with unit spectral norm,
% of the corresponding dimension, 
all elements of $\bE_{t,i,j}$ are independent and $N(0,1)$. Matrices $\bPhi_{1,i,j}$ and $\bPhi_{2,i,j}$ are constructed such that their singular values are all one and the left and right singular matrices are orthonormal, and the scalar coefficient $\lambda_{i,j}$ controls the level of auto-correlation and its values for the four blocks are marked in Table~\ref{table:example3}.

%\begin{align*}
%\bPhi_{1,ij}&:=
%\begin{array}{c|c|c|c|c|c}
% & j=1 & 2 & 3 & \ldots & n_r \\ \hline
%i=1& 0.9 & 0.8 & [0.1,0.3]& \ldots & [0.1,0.3] \\ \hline 
%2& 0.5 & 0.5 & 0.5& \ldots & 0.5 \\ \hline 
%3& 1 & 1 & 1 & \ldots & 1 \\ \hline 
%\vdots & \vdots & \vdots & \vdots & \ddots & \vdots \\ \hline
%n_r& 1 & 1 & 1 & \ldots & 1 \\ \hline 
%\end{array} 
%\end{align*}
%\begin{align*}
%\bPhi_{2,ij}&:=
%\begin{array}{c|c|c|c|c|c}
% & j=1 & 2 & 3 & \ldots & n_c \\ \hline
%i=1& 0.9 & 0.8 & 1& \ldots & 1 \\ \hline 
%2& 0.5 & 0.5 & 1& \ldots & 1 \\ \hline 
%3& [0.1,0.3] & [0.1,0.3] & [0.1,0.3] & \ldots & [0.1,0.3] \\ \hline 
%\vdots & \vdots & \vdots & \vdots & \ddots & \vdots \\ \hline
%n_c& [0.1,0.3] & [0.1,0.3] & [0.1,0.3] & \ldots & [0.1,0.3] \\ \hline 
%\end{array}
%\end{align*}
\end{example}

We set sample sizes $T=500,1000,2000,5000$, and $(p,q)=(6,6),(8,6),(8,8),(10,10)$. For each setting, we simulate a matrix time series of length $T+M$ with $M=10$. To perform one-step ahead post-sample forecasting, for each $i=0, 1, \cdots, M-1$, we use the first $T$ observations for identifying the uncorrelated blocks and for computing $\wh \bA_*, \wh \bB_*$ and $\wh \bU_t^*$ defined in (\ref{est-3}). Depending on its shape/size, we fit each identified block in $\wh \bU_t^*$ with an MAR(1), VAR(1) or AR(1) model. The fitted models are used to predict $\bU_{T+i+1}^*$. The predicted value for $ X_{T+i+1}$ is then obtained by
\begin{align}\label{model2:ex3}
\wh X_{T+i+1}= \big(\wh\bSigma^{(x)}_2\big)^{1/2}
\wh\bB_*\wh\bU_{T+i+1}^* \wh\bA_*^\top \big(\wh\bSigma^{(x)}_1\big)^{1/2}, \quad i=0, 1,
\cdots, M-1.
\end{align}
We then compute the mean squared error: 
\begin{equation} \label{rmse}
{\rm MSE}=\frac{1}{pq}\sum_{i=1}^p\sum_{j=1}^q
(\wh X_{s,i,j}-\mu_{s,i,j})^2,
\end{equation}
where $\wh X_{s,i,j}$ denote the one-step ahead predicted value for the $(i,j)$-th element $X_{s,i,j}$ of $ X_s$, and $\mu_{s,i,j} =
\E(X_{s,i,j}| X_{s-1}, X_{s-2}, \cdots)$. We compare $\wh X_{s,i,j}$ with $\mu_{s,i,j}$, instead of $ X_{s,i,j}$, to remove the impact of the noise in the observed $ X_{s,i,j}$. We report overall the mean and the standard deviation of MSE over 1000 replications.
%In addition we also report the mean and the standard deviation over the replications with the completely correct segmentation of the uncorrelated blocks in $\bU_t^*$, over those with nearly correct segmentation (i.e. either $\wh n_r = n_r -1$ or $\wh n_c = n_c -1$), and over all the other cases.

For the comparison purpose, we also include several other models in our simulation: 
(i) VAR(1) model, i.e. $ X_t$ is stacked into a vector of length $pq$ and a VAR(1) is fitted to the vector series directly, (ii) MAR(1) model, i.e. MAR(1) of \cite{chen2021autoregressive}, is fitted directly to $ X_t$, and (iii) TS-PCA, i.e. all the elements of $ X_t$ are stacked into a long vector and the segmentation method of \cite{chang2018} is applied to the vector series.
In addition, we also include two oracle models:
% for the evaluation of the sensitivity of different component of the segmentation procedure:
(O1) the true segmentation blocks are used to model $\wh\bU_t^*$; and
%(O2) true $\Sigma^{(x)}_1,\Sigma^{(x)}_2$ are used for normalization in the first step and the true segmentation blocks are used to model $\wh\bU_t^*$; and
(O2) True $\Sigma^{(x)}_1,\Sigma^{(x)}_2$ and true $\bA_*,\bB_*$ are used and the true segmentation blocks are used to model $\wh\bU_t^*$.

%\begin{enumerate}
%\item[O1.] the model based on the true segmentation groups;
%\item[O2.] the model based on the true segmentation groups and using true $\Sigma^{(x)}_1,\Sigma^{(x)}_2$;
%\item[O3.] the model based on the true segmentation groups and using true $\Sigma^{(x)}_1,\Sigma^{(x)}_2$ and $A_1,B_1$ (cf. \eqref{a4}).
%\end{enumerate}
% in addition to show the overall prediction performance of the segmentation procedure,
% we also compute the RMSE for show the percentage of correct block identification, percentage of correct block identification except one (and only one) pair of column blocks or row blocks being merged (near correct block identification), and the percentage of all other incorrect identifications. 
%mean and standard deviations under correct segmentation case, incomplete segmentation case ($(\wh n_r,\wh n_c)=(n_r-1,n_c-1), (n_r-1,n_c)$ or $(n_r,n_c-1)$) and other specification cases, and report the corresponding proportions. 

The mean and standard deviations (in bracket) of the MSEs are listed in Table \ref{tab3}. It is clear that the forecasting based on the proposed decorrelation transformation is more accurate than those based on MAR(1), VAR(1) and TS-PCA directly. The improvement is often substantial.
The only exception is the case of $(p,q)=(6,6)$ and $T=5000$: the 36-dimension VAR(1) performs marginally better (i.e. the decrease of 0.002 in MSE) than the transformation based method. On the other hand, the improvement of the transformation based forecast over VAR(1) for sample size $T \le 2000$ is more significant. The relative poor performance of MAR(1) is due to the substantial discrepancy between MAR(1) model and the data generating mechanism used in this example. Note that there are $(p+q)^2$ free parameters in VAR(1)  while there are merely $(p^2+q^2)$ parameters in MAR(1). With large $T$, VAR(1) is more flexible than MAR(1). TS-PCA performs poorly, due to the {large} error in
estimating the large transformation matrix of size $pq\times pq$.
%depends on the total dimension $pq$ of the vectorized series, which is large even for moderate $p$ and $q$; 
See also Remark \ref{rm-rates}. In addition, TS-PCA also requires to fit VAR models for several large blocks of size 9, 6, 6 and etc.
%In addition, 
%there are a lot of \rc{The} large blocks (of size 9, 6, 6, 4, etc), further increases the difficulty of correctly identifying uncorrelated blocks \rc{using TS-PCA}.}

%Table~\ref{tab3} indicates that when one pair of blocks is merged, the prediction performance remain relatively stable, comparing to the correct block identifications. This is partly due to that, most of such mergers are combination of two univariate blocks. The result is to use a two-dimensional VAR(1) model instead of two one-dimensional AR(1) models. Hence when sample size is relatively large, the minor overfitting does not cause major performance deterioration. Overall, when segmentation is correct or mostly correct, the prediction performance is the best. When sample size is small, the incorrect segmentation still provide better prediction performance. %As expected, all models deteriorate as $(p,q)$ increases. 
%{\color{red} The percentages for "others" are very high. It also conveys a positive message: even the segmentation is poor, the prediction is still better. But readers may be worried, as if the method does not work well even with T=5K. I suggest to report the segmentation results only, not the 3 divided cases -- QY.}

Table \ref{tab3} indicates that the oracle model O1 performs only slightly better than segmentation when sample size $T=500$ or 1000. %, and is actually worse than the cases when the segmentation produces correct or nearly correct block identifications. 
Oracle Model O2 only deviates from the true model in terms of the estimation error of the coefficients in the AR models, hence performs the best. 
% that both correct segmentation and incomplete segmentation lead to more accurate forecasts than  oracle model O1 and O2, and oracle model O1 and O2 are better than other segmentation cases. Together, the proposed method is worse than O1 and O2. 
%Comparing the performance of O1 and O2, it seems that the estimation of $\Sigma^{(x)}_1$ and $\Sigma^{(x)}_2$ does not have large impact on the prediction performance. 
%\yfm{Since MSE in \eqref{rmse} \rc{does not include the noise in the observations to be predicted}, MSE of O2 is very close to 0. When $T$ is large, the VAR(1), proposed segmentation and O1 \rc{have} similar performance that is much worse than O2. It shows that even for large $T$, the estimation error of the inverse of sample covariance matrix in the proposed segmentation, O1 or the least square estimator of VAR(1) is still not negligible. } 
The MSE of O2 is very close to 0, since the only source of error in MSE is the estimation error of the AR models for each blocks of $U_t$, which goes to zero quickly as $T$ increases since these models are of small dimensions. When $T$ is large, the VAR(1), proposed segmentation and O1 all perform about the same, which is worse than O2. 

\begin{table}[htbp]
\caption{Example 3 -- One-step ahead post-sample forecasting: means and standard deviations (in bracket) of MSEs in a simulation with 1000 replications.}\label{tab3}
%\vspace{-7mm}
%\small{
\scriptsize{
\begin{center}
\begin{tabular}{l|l|ccccccccccc}
\multicolumn{2}{c|}{} & \multicolumn{8}{c}{MSEs and standard deviations} \\
\hline
\multicolumn{2}{c|}{} & \multicolumn{2}{c}{$T=500$} & \multicolumn{2}{c}{$T=1000$} & \multicolumn{2}{c}{$T=2000$} & \multicolumn{2}{c}{$T=5000$} \\
 \hline \hline
\multirow{6}{*}{$(p,q)$= $(6,6)$ } & MAR(1) &  0.515(0.333) & & 0.501(0.304) & & 0.513(0.305) & & 0.482(0.280) & \\
& VAR(1) & 0.369(0.142) & & 0.231(0.065) & & 0.137(0.037) & & 0.044(0.014)  \\ 
& TS-PCA & 0.492(0.232) & & 0.278(0.094) & & 0.149(0.043) & & 0.048(0.018)\\
& segmentation & 0.257(0.180) & & 0.144(0.098) & & 0.087(0.094) & & 0.046(0.051)  \\ \cline{2-10}
& model O1  &  0.207(0.173) & & 0.098(0.079) & & 0.046(0.033) & & 0.018(0.012)   \\
& model O2  &  0.018(0.010) & & 0.009(0.005) & & 0.004(0.003) & & 0.002(0.001)  \\
 \hline \hline
\multirow{6}{*}{$(p,q)$= $(6,8)$ } & MAR(1) & 0.981(0.496) & & 0.977(0.491) & & 0.947(0.488) & & 0.942(0.485) & \\
& VAR(1) & 0.549(0.235) & & 0.323(0.111) & & 0.179(0.058) & & 0.095(0.022) \\ 
& TS-PCA & 0.787(0.359) & & 0.442(0.168) & & 0.197(0.068) & & 0.096(0.023) \\
& segmentation & 0.329(0.262) & & 0.166(0.125) & & 0.097(0.073) & & 0.061(0.049)  \\ \cline{2-10}
& model O1  & 0.233(0.120) & & 0.127(0.064) & & 0.078(0.041) & & 0.051(0.035)  \\
& model O2  & 0.032(0.015) & & 0.016(0.007) & & 0.008(0.004) & & 0.003(0.001) \\
 \hline \hline
\multirow{6}{*}{$(p,q)$= $(8,8)$ } & MAR(1) & 0.971(0.434) & & 0.924(0.363) & & 0.915(0.391) & & 0.908(0.384) & \\
& VAR(1) & 0.861(0.304) & & 0.447(0.148) & & 0.275(0.079) & & 0.121(0.033) & \\ 
& TS-PCA & 0.965(0.420) & & 0.564(0.215) & & 0.290(0.106) & & 0.129(0.036)\\
%& \yfm{permutationMax} & \yfm{1.978} & & \yfm{1.586} & & \yfm{1.281} & & \yfm{0.993} \\
%& \yfm{permutationFDR (0.001)} & \yfm{1.147} & & \yfm{0.657} & & \yfm{0.386} & & \yfm{0.186} \\
%& \yfm{permutationFDR (0.01)} & \yfm{1.066} & & \yfm{0.618} & & \yfm{0.347} & & \yfm{0.149} \\
%& \yfm{permutationFDR (0.05)} & \yfm{0.965} & & \yfm{0.553} & & \yfm{0.290} & & \yfm{0.128} \\
%&\yfg{use true $r=96$} & \yfg{0.914} & & \yfg{0.531} & & \yfg{0.388} & & \yfg{0.197} \\
%&\yfg{fix $\hat r=(2/3)r=64$} & \yfg{1.025} & & \yfg{0.630} & & \yfg{0.449} & & \yfg{0.249} \\
%&\yfg{fix $\hat r=120$} & \yfg{0.872} & & \yfg{0.479} & & \yfg{0.333} & & \yfg{0.166} \\
& segmentation & 0.499(0.276) & & 0.271(0.158) & & 0.176(0.131) & & 0.091(0.092)  \\ \cline{2-10}
& model O1  & 0.399(0.189) & & 0.252(0.154) & & 0.193(0.137) & & 0.120(0.133)  \\
& model O2  & 0.040(0.015) & & 0.019(0.007) & & 0.010(0.004) & & 0.004(0.001) \\
 \hline \hline
\multirow{6}{*}{$(p,q)$= $(10,10)$ } & MAR(1) & 1.244(0.512) & & 1.214(0.521) & & 1.207(0.481) & & 1.218(0.509) & \\
& VAR(1) & 1.920(0.654) & & 1.151(0.353) & & 0.646(0.191) & & 0.354(0.078) \\ 
& TS-PCA & 1.822(0.711) & & 1.216(0.413) & & 0.726(0.211) & & 0.365(0.128) \\
& segmentation & 0.967(0.505) & & 0.664(0.326) & & 0.465(0.240) & & 0.312(0.224)  \\ \cline{2-10}
& model O1  & 0.852(0.403) & & 0.617(0.307) & & 0.446(0.235) & & 0.341(0.233)  \\
& model O2  & 0.062(0.022) & & 0.030(0.010) & & 0.016(0.005) & & 0.006(0.002)  \\
 \hline \hline
\end{tabular}
\end{center}
}
\end{table}

\begin{example}
Next, we assess the sensitivity of the proposed matrix segmentation method with respective to the Kronecker product structure ($\vec(X_t)=(A\otimes B) \vec(U_t)$). We use the same model setting in Example 3 and choose $(p,q)=(8,8), T=500$. Instead of model \eqref{a1}, we consider $\vec(X_t)=(A\otimes B+\Psi) \vec(U_t)$, where $\Psi$ is a perturbation matrix with $N$ none-zero elements. We also set each non-zero element of $\Psi$ to be $c \|A\otimes B\|_F/(pq)\times \omega$, $\omega\overset{\rm iid}{\sim}$ Rademacher distribution. Table \ref{tab4} shows MSEs and standard deviations of VAR(1) and the proposed matrix segmentation method over 1000 replications, for several combinations of $N$ and $c$. Again, $N$ controls the number of non-zero perturbations and $c$ controls the level of the perturbations, comparing to the average size of the elements in $A\otimes B$. When $c=0$, it becomes the original case in Table \ref{tab3}. From Table~\ref{tab4}, it is seen that the segmentation is still useful in prediction when the model is deviated from the assumed block structure within certain perturbation. The prediction performance of the VAR model is not affected by the perturbation since it does not assume any block structure. It is seen that the segmentation outperforms the VAR model when $c<0.3$ when $N=10$ and $c<0.2$ when $N=100$. This feature demonstrates the benefit of dimension reduction through segmentation, even when the underlying model does not have the exact Kronecker product structure. 
%The performance of the TS-PCA approach does not change under perturbation because a dominating portion of the prediction error comes from the estimation error of the high dimensional transformation matrix.} 
%the proposed matrix segmentation method has smaller MSEs than VAR(1) and TS-PCA in all cases except $N=100,c=0.3$. This indicates that our segmentation method is superior than other competing methods for post-sample forecasting, even under moderate violation of the Kronecker product structure $A\otimes B$.
%As expected, the performances of VAR(1) and TS-PCA are relatively stable to $\Psi$.
\end{example}

\begin{table}[htbp]
\caption{Sensitivity of the Kronecker product structure using different $N$ and $c$. The results are the means and standard deviations (in bracket) of MSEs of one-step ahead post-sample forecasting based on 1000 replications.}
\label{tab4}
%\vspace{-7mm}
%\small{
\scriptsize{
\begin{center}
\begin{tabular}{l|l|ccccccccccc}
\multicolumn{2}{c|}{} & \multicolumn{5}{c}{MSEs and standard deviations} \\ \hline
\multicolumn{2}{c|}{} &  $c=0$ & $c=0.05$ & $c=0.15$ & $c=0.2$ & $c=0.3$ &\\
 \hline \hline
\multirow{3}{*}{$N=10$ } & VAR(1) & 0.861(0.304) &  0.877(0.327) &  0.872(0.314) &  0.874(0.291) &  0.869(0.306) \\
%& TS-PCA & 0.965(0.420) &  0.969(0.404) &  0.959(0.439) &  0.967(0.421) &  0.968(0.420) \\
& segmentation & 0.499(0.276) &  0.490(0.241) &  0.620(0.303) &  0.732(0.399) &  0.859(0.504) \\ 
 \hline \hline
\multirow{3}{*}{$N=100$ } & VAR(1) & 0.861(0.304) &  0.856(0.295) &  0.860(0.303) & 0.874(0.326) & 0.878(0.317) \\
%& TS-PCA & 0.965(0.420) & 0.974(0.431) & 0.941(0.402) & 0.970(0.434) & 0.983(0.467)\\
& segmentation & 0.499(0.276) & 0.570(0.314) & 0.721(0.458) & 0.856(0.530) & 0.976(0.689) \\ 
 \hline \hline
\end{tabular}
\end{center}
}
\end{table}

\subsection{Real data analysis} \label{section:data}

We now illustrate the proposed decorrelation method with a real data example. The data concerned is a 17$\times$6 matrix time series consisting of
the logarithmic daily averages of the 6 air pollutant concentration readings (i.e. PM$_{2.5}$, PM$_{10}$, SO$_2$, NO$_2$, CO, O$_3$) from the 17 monitoring stations in Beijing and its surrounding areas (i.e. Tianjin and Hebei Province) in China.
The sampling period is January 1, 2015 -- December 31, 2016 for a total of $T=731$ days. %The PM$_{2.5}$ series was analyzed in \citet{chang2018}.
%to demonstrate the benefit of time series principal component analysis (TS-PCA). 
The readings of different pollutants at different locations are crossly correlated. %Figure \ref{figure:row} shows the cross correlogram of five randomly selected stations for PM$_{10}$, and Figure \ref{figure:col} displays the cross correlogram of all six pollutants for a station in Beijing. 

We first remove the seasonal mean of the $17\times 6$ matrix time series. Setting $\tau_0=5$ in (\ref{est-2}), we apply the proposed bilinear transformation to discover the uncorrelated block structure.
The finding is stable:
with $5\le\tau_1\le 30$ in (\ref{a13}), the transformed matrix series admits $\wh n_c=4$ uncorrelated column blocks with sizes $3,1,1, 1$ respectively, and $\wh n_r=15$ uncorrelated row blocks with two blocks of size 2 and the rest of size 1. Overall there are $15\times 4$ blocks, among which there are 2 blocks of size $2\times 3$, 13 blocks of size $1\times 3$, 6 blocks of size $2\times 1$ and 39 blocks of size $1\times 1$.
% Note that the blocks do not have clear interpretations as the original matrix time series has been transformed. 

To check the post-sample forecasting performance, we calculate the rolling one-step, and two-step ahead {out-sample} forecasts for each of the daily readings in the last three months in 2016 (total 92 days). The methods included in the comparison are (i) the forecasting based on the segmentation derived from the proposed decorrelation transformation by fitting an MAR(1), VAR(1) or AR(1) to each identified block according to its size and shape; (ii) MAR(1) for the original $17\times 6$ matrix series; (iii) VAR(1) for the vectorized original series, (iv) univariate AR(1) for each of the $17\times 6$ original series, and (v) the TS-PCA \citep{chang2018} for the vectorized original series. The TS-PCA leads to the segmentation consisting of one group of size 3, three groups of size 2, and 93 groups of size 1. {For each group, a VAR(1) model is used for prediction.}

For the two segmentation approaches, we fix the needed transformation obtained using data up to September 30, 2016, and the corresponding segmentation structure. The time series model for each individual block is updated throughout the rolling forecasting period.

In addition to the identified segmentation with 15$\times$4 blocks, we also compute the forecasts based on two other segmentations: one with 14$\times$3 blocks, and another with 16$\times$5 blocks. The former is obtained by merging two most correlated single row blocks in the discovered segmentation into one block and two most correlated single column blocks into one block. The latter is obtained by splitting one of the two blocks with two rows in the discovered segmentation into two single row blocks and splitting the block with 3 columns into two blocks. This will reveal the impact of slightly `wrong' segmentations on forecasting performance.
%three block-structures (i) $15\times 4$ groups with partition $(2,2,1,\ldots,1)\times (3,1,1,1)$; (ii) $14\times 3$ groups, with a merging of two closest one-dimensional row blocks, and two closest one-dimensional column blocks; and (iii) $16\times 5$ groups, with a slitting of the weaker 2-dimensional row block into two 1-dimensional row blocks, and the 3-dimensional column block into a one 2-dimensional column block and one 1-dimensional column block. 

% Under each method, we report 
% the mean squared predictive errors (MSPE) for one-step ahead forecasting,
% $$\frac{1}{h_0} \sum_{h=1}^{h_0} (\wh  X_{T_0+h,ij} -  X_{T_0+h,ij})^2 ,$$
% where $\wh  X_{T_0+h}$ is forecasting of $ X_{T_0+h}$ given all data $ X_1,\ldots, X_{T_0+h-1}$, where $h_0=92$ and $T_0=639$ in this excise. 

\begin{table}[ht]
\scriptsize{
\caption{One-step and two-step ahead post-sample forecasting for the air pollution data: means and standard deviations (in bracket) of MSPE$_s$, regrets and adjusted ratios of the various methods.}
% in sample variance of VAR(1) is 0.197, VAR(2) is 0.164}}
\label{tab:pollution}
\ \\
\centering
\begin{tabular}{l|ccc|ccccccc}\hline
\multirow{2}{*}{Method} & \multicolumn{3}{c|}{One-step forecast}  &\multicolumn{3}{c}{Two-step forecast} \\ \cline{2-7}
& raw forecast & regret & adjust ratio & raw forecast & regret & adjust ratio \\\hline
MAR(1) & 0.351 (0.208) & 0.154&1.305 & 0.516 (0.303) & 0.319 & 1.340 \\
VAR(1) & 0.400 (0.272) & 0.203&1.720 & 0.717 (0.406) & 0.520 & 2.185 \\
Univariate AR & 0.346 (0.205) & 0.149&1.263 & 0.501 (0.286) & 0.304 & 1.277 \\ 
TS-PCA & 0.332 (0.188) & 0.135&1.144 & 0.450 (0.247) & 0.253 &  1.063  \\ \hline
Segmentation with 15$\times$4 blocks & 0.315 (0.182) & 0.118 &1.000 & 0.435 (0.246) & 0.238 & 1.000 \\
Segmentation with 14$\times$3 blocks & 0.319 (0.185) & 0.122 &1.034 & 0.446 (0.256) & 0.249 & 1.046 \\
Segmentation with 16$\times$5 blocks & 0.318 (0.183) & 0.121 &1.025 & 0.443 (0.247) & 0.246 & 1.034 \\
\hline
\end{tabular} }
\end{table}

Let the mean squared forecast error for $s$-th observation be
\begin{equation} \label{mspe}
{\rm MSPE}_s=\frac{1}{pq}\sum_{i=1}^p\sum_{j=1}^q
(\wh X_{s,i,j}-X_{s,i,j})^2,
\end{equation}
where $\wh X_{s,i,j}$ denote the one-step ahead predicted value for the $(i,j)$-th element $X_{s,i,j}$ of $ X_s$. Two-step ahead MSPE$_s$ is defined similarly. The means and the standard deviations of MSPE$_s$ of one-step and two-step ahead post-sample forecasts for the pollution readings in the last 92 days are listed in Table \ref{tab:pollution}. The prediction based on the bilinear decorrelation transformation (even using `wrong' segmentations) is clearly more accurate than those without transformation as well as that of the TS-PCA. Using the `wrong' segmentation deteriorates the performance only slightly, indicating that a partial (instead of total) decorrelation still leads to significant gain in prediction. Applying TS-PCA to the vectorized series requires to estimate a $102\times 102$ transformation matrix (instead of the $17\times 17$ and $6\times 6$ matrices {by utilizing the matrix time series structure}). It leads to larger estimation errors (see Remark \ref{rm-rates} in Section \ref{section:theories}). Nevertheless it still provides more accurate predicts than those without transformation. 
Also note that fitting each of the original $17\times 6$ time series with a univariate AR(1) separately leads to a better performance
than those from fitting MAR(1) and VAR(1) to the original matrix series jointly. This indicates that the cross-serial correlation is useful and important information for future forecasting. However to make efficient use of the information, it is necessary to adopt some effective independent component analysis or dimension-reduction techniques such as the proposed decorrelation transformation which pushes correlations across different series into the autocorrelations of some transformed series.
% \rc{On the other hand, applying the TS-PCA procedure \cite{chang2018} on the 104 dimension vectorized series results in a more aggressive segmentation structure with one group of size $3$, three groups of size $2$ and $95$ groups of size $1$, using an estimated transformation matrix of size $107\times 107$. Its out-sample prediction performance is not as good as the proposed segmentation directly on the matrix time series, with much smaller transformation matrices. It may be due to over-estimation, as the sample size is $T=639$ for the training period.} 

%\rc{%The apparent small improvement in Table 5 is partially due to the inherent noise in the observed $X_{t+1}$ that are used in calculating the prediction performance and are shared by all prediction methods. In this revision we added additional columns in Table 5, showing the `regret' of using various prediction methods against a `linear Oracle' that uses full dataset to form a linear predictor for prediction. 
Also included in the table are `regret' defined as the difference between MSPE and the in-sample residual variance of  the fitted VAR(1) model for the vectorized data  (i.e. a vector time series of dimension $17\times 6= 102$), and `adjusted ratio' defined as the ratio of the regret to the regret of the best prediction model (i.e. the segmentation with $15\times 4$ blocks). The regret tries to measure the forecasting error on the predictable signal, removing the impact from unpredictable noise in the model. Since the fitted VAR(1) model uses more than 10,000 parameters,
its sample residual variance underestimates the noise variance in the model, and is taken as a proxy for the latter.
%The additional `regret' columns shown in the table are the difference between the estimated in-sample residual variance from the linear Oracle and the prediction MSE of various models. 
%The linear Oracle we used is a full VAR(1) model on the vectorized time series and its estimated in-sample residual variance is used as the baseline unavoidable prediction error. As the VAR(1) model uses more than 10000 parameters in the AR coefficient matrices, with total 700 observations of size $17\times 6$, the model significantly overfits the data, hence the in-sample residual variance tends to underestimate the true noise variance, the `regret' shown is a conservative one.} The columns of `adjusted ratio' are calculated by subtracting the in-sample error variance (of the best model) from each prediction error, and using the $15\times 4$ segmentation result as the base line to calculate the ratio. From the extra columns, 
The adjusted ratios shows that the `regret' of TS-PCA is 14.4\% worse than the $15\times 4$ segmentation for one-step ahead forecast, and 6.4\% worse than the $15\times 4$ segmentation for two-step ahead forecast.

\begin{figure}[ht]
\centering
\includegraphics[scale=0.4,page=1]{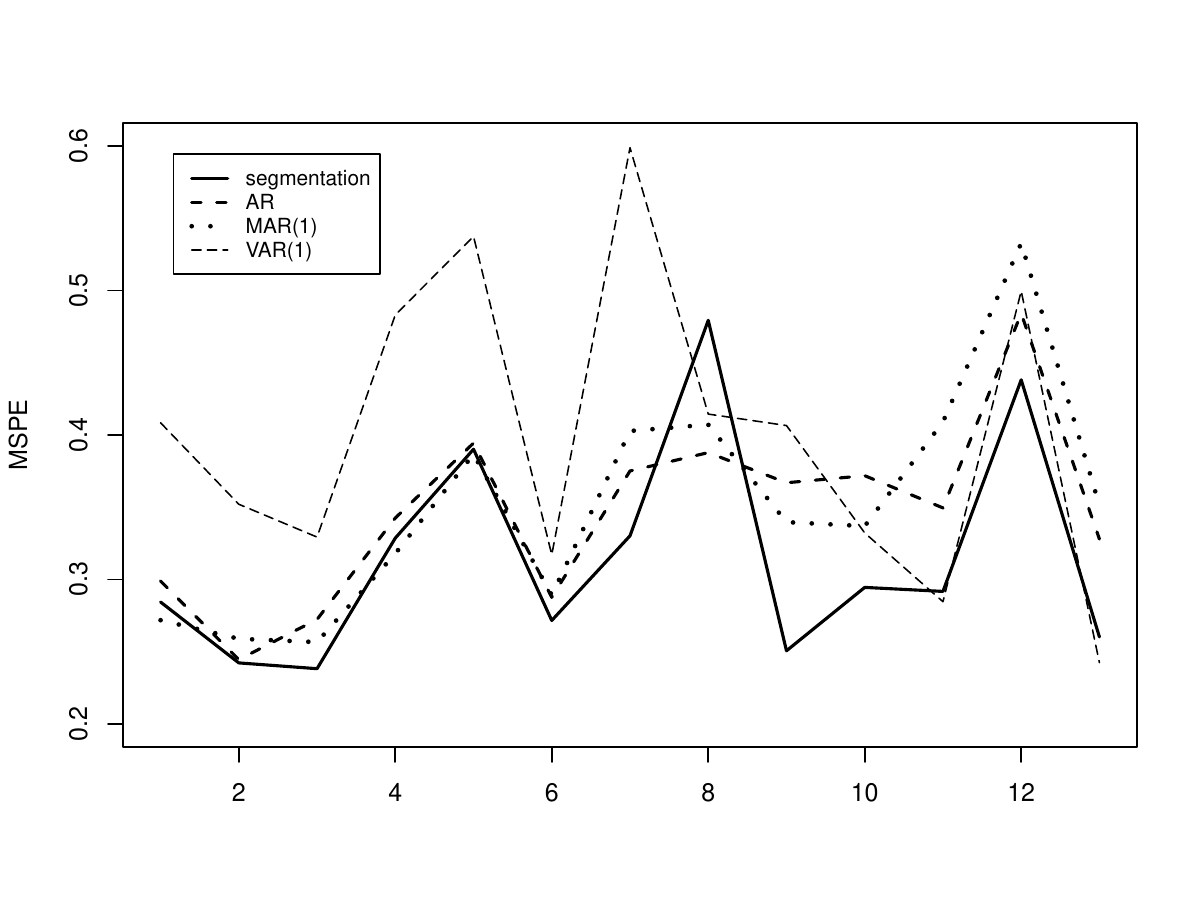}
\caption{Weakly averaged one-step forecast errors of the univariate AR(1) models for each component series (AR), the matrix AR(1) model (MAR(1)), the vector AR(1) for vectorized series (VAR(1)), the method based on
the bilinear decorrelation transformation (segmentation).}
\label{figure:weakly_mean}
\end{figure}

To further evaluate the forecasting stability, Figure \ref{figure:weakly_mean} shows weekly average of one-step rolling forecast MSPE within the forecasting period under various models. It is seen that the proposed segmentation outperforms the other three methods in most of the periods in terms of prediction.

%we calculate the rolling MSPE for one-step ahead post-sample forecasts based on, respectively, the segmentation, MAR(1), VAR(1) and univariate AR models. We plot the weekly averages of the rolling MSPE in Figure \ref{figure:weakly_mean}. It shows that the forecast based on the proposed segmentation almost always outperforms the other three methods.

%The proposed method identifies a segmentation with
In this example, the identified segmentation of $(\wh n_r, \wh n_c)= (15,4)$ for the original 17$\times$6 matrix series is more likely to be an approximation {of} the underlying dependence structure rather than a true model. The fact that the two `wrong' segmentation models (though quite close to the discovered one) as well as the aggressive segmentation via vectorized TS-PCA transformation provide comparable, though inferior, post-sample forecasting performance lends further support to the claim that the proposed decorrelation method makes the transformed series more predictable than the original ones, regardless model (\ref{a1}) holds or not. See Remark \ref{rm-model}(iv) in Section \ref{sec21} above.

%\begin{figure}[htbp]
%\centering
%\includegraphics[scale=0.6,page=1]{rolling_mean}
%\caption{The rolling means of one-step ahead post-sample forecasting for 30 days and averaging over all $p,q$ series}
%\label{figure:pollution}
%\end{figure}

\bibliographystyle{apalike}
%\bibliography{multiple-iso.bib}
\bibliography{simultaneous-decomp}

%\end{document}
\newpage

\appendix
\setcounter{page}{1}

\begin{center}
{\LARGE\bf Supplementary Material to ``Simultaneous Decorrelation of Matrix Time Series''}

\author{Yuefeng Han, Rong Chen, Cun-Hui Zhang and Qiwei Yao \\
University of Notre Dame, Rutgers University and London School of Economics}
\end{center}

\begin{abstract}
In the supplementary material, we provide additional simulation results (Section \ref{section:add_sim}), additional information on the air pollutant example (Section \ref{section:plots}) and all proofs (Section \ref{section:appendix}).
\end{abstract}

\section{Additional simulation results}\label{section:add_sim}
%\noindent \bf Example 4:

\begin{figure}[htb!]
\centering
\subfigure[Effect of $\tau_0$ (fix $\tau_1=15$)]{
\includegraphics[scale=0.36]{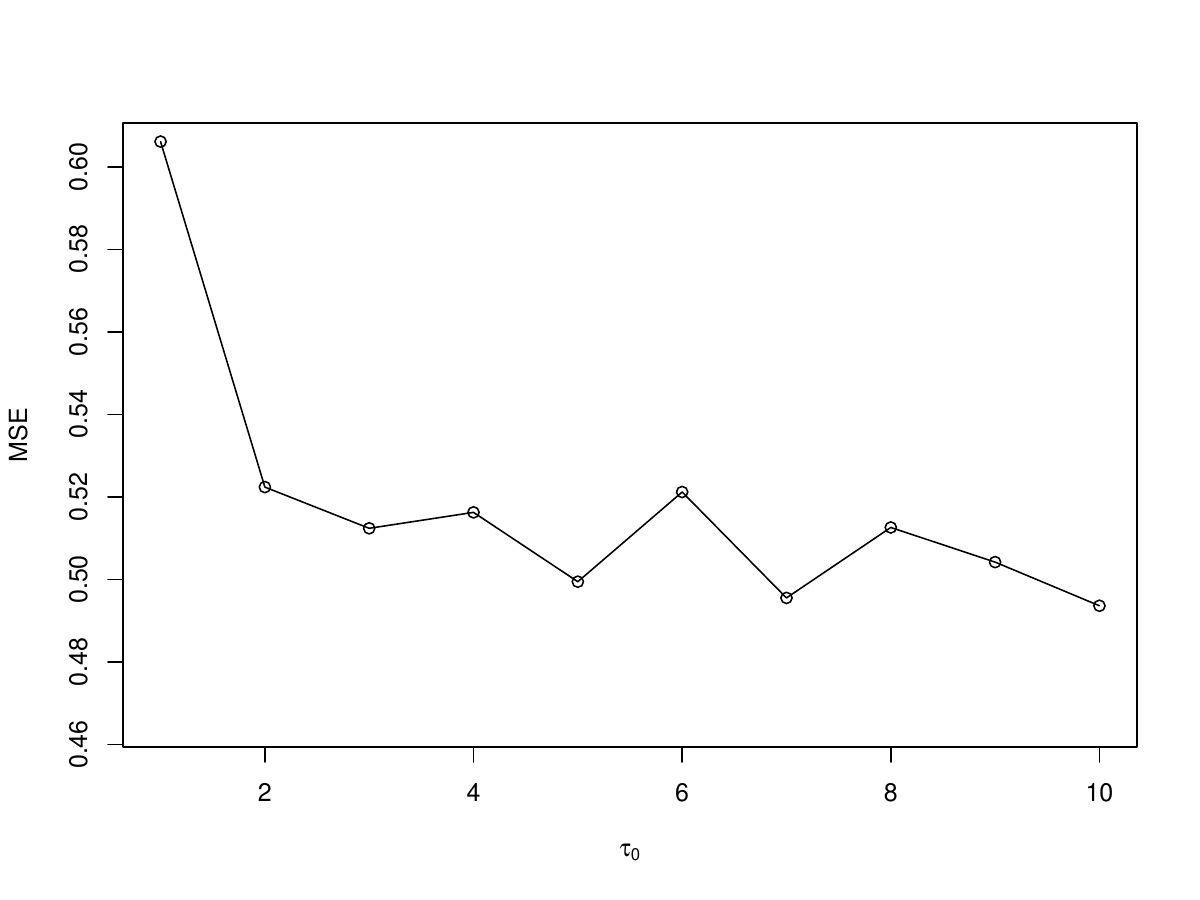}}
\subfigure[Effect of $\tau_1$ (fix $\tau_0=5$)]{
\includegraphics[scale=0.36]{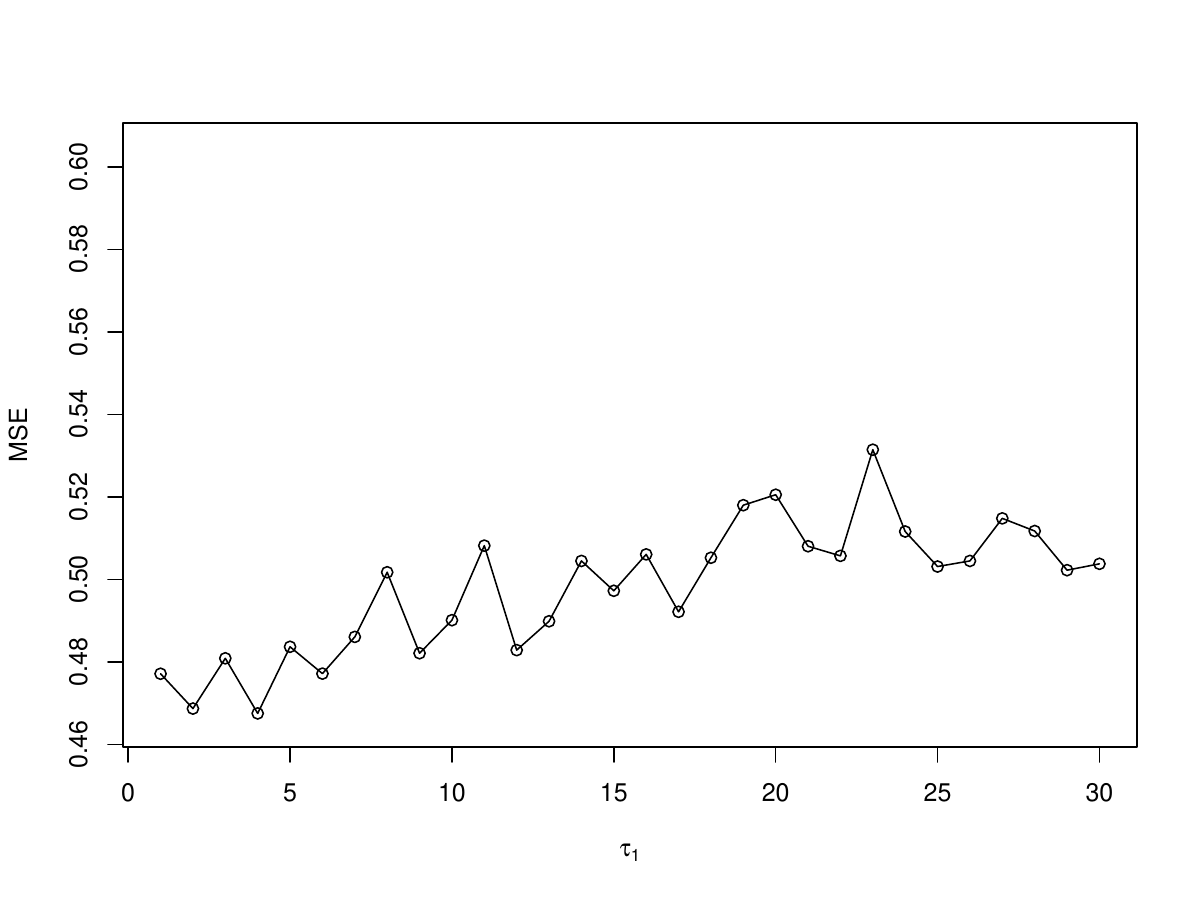}}
\caption{Sensitivity analysis with changing $\tau_0$ and $\tau_1$}
\label{figure:tau}
\end{figure}

%\noindent \bf Example 5:
\begin{example}
In this example we provide some sensitivity analysis with the tuning parameters $\tau_0$ in \eqref{aa9} and $\tau_1$ in \eqref{a13}, as discussed in Remark~\ref{remark:corPairs}(ii). Again, we use the same model setting in Example 3 and choose $(p,q)=(8,8), T=500$. Note that $\tau_0=5$ and $\tau_1=15$ are used in Example 3. Figure \ref{figure:tau} shows the average of post sample MSEs over 1000 independent Monte Carlo experiments for different combinations of $\tau_0$ and $\tau_1$. It can be seen from Figure \ref{figure:tau}(a) that except $\tau_0=1$, the segmentation performance remains relatively stable for different values of $\tau_0$.
%MSE remains almost the same, and fluctuates among 0.50 to 0.52. 
Similarly, Figure \ref{figure:tau}(b) shows that the segmentation performance using different $\tau_1$ is also relatively stable. Since the data generating process is an AR process of order 1, the cross-autocorrelations are the largest at the small lags, hence only a relatively small $\tau_1$ is needed in this case. Overall, the MSE does not change much with $\tau_0,\tau_1$.
\end{example}

\section{Additional information on the air pollutant example}\label{section:plots}

\begin{figure}[htbp]
\centering
\includegraphics[scale=0.55,page=1]{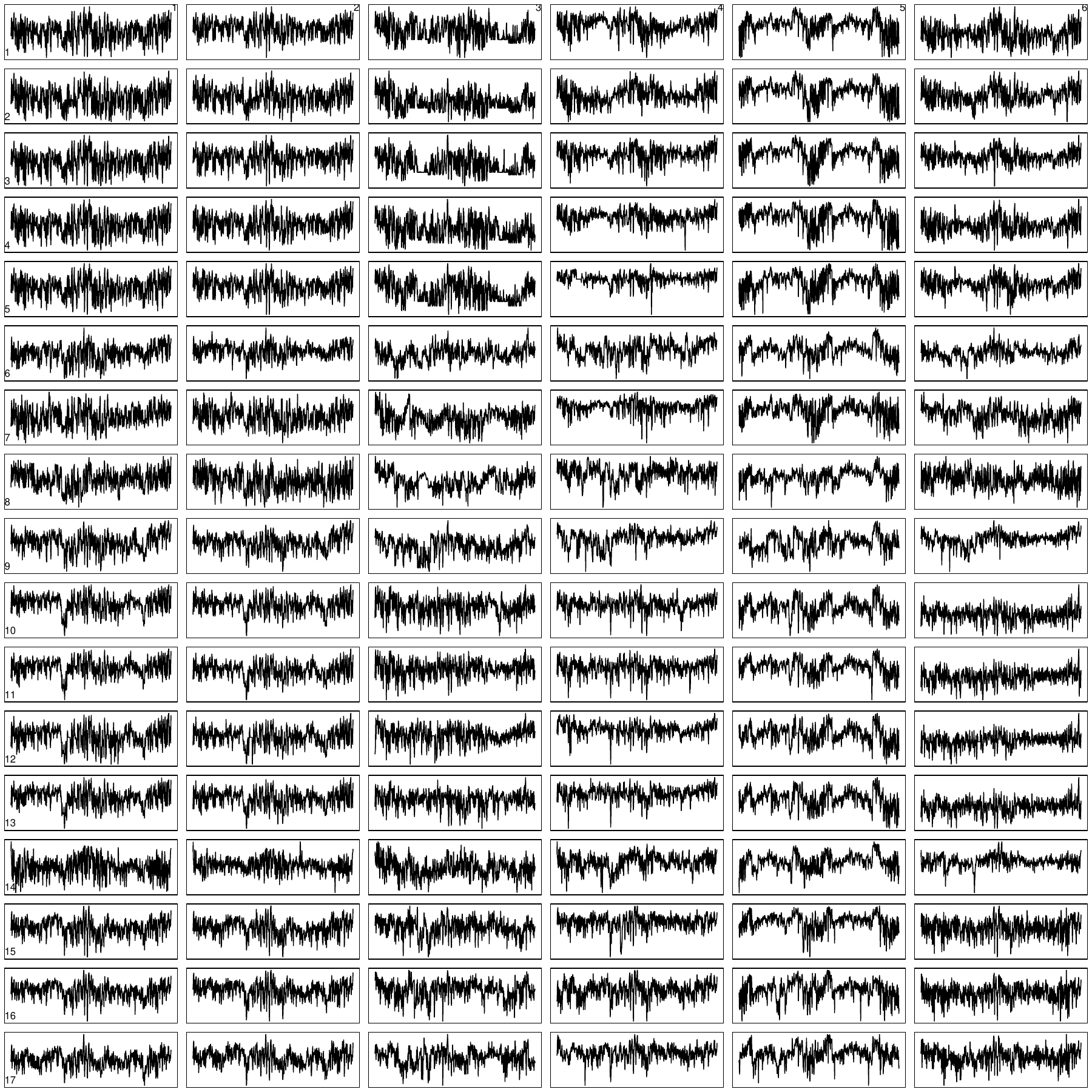}
\caption{Time series plot of 17 by 6 air pollution data}
\label{figure:data}
\end{figure}

\begin{figure}[htbp]
\centering
\includegraphics[scale=0.5,page=1]{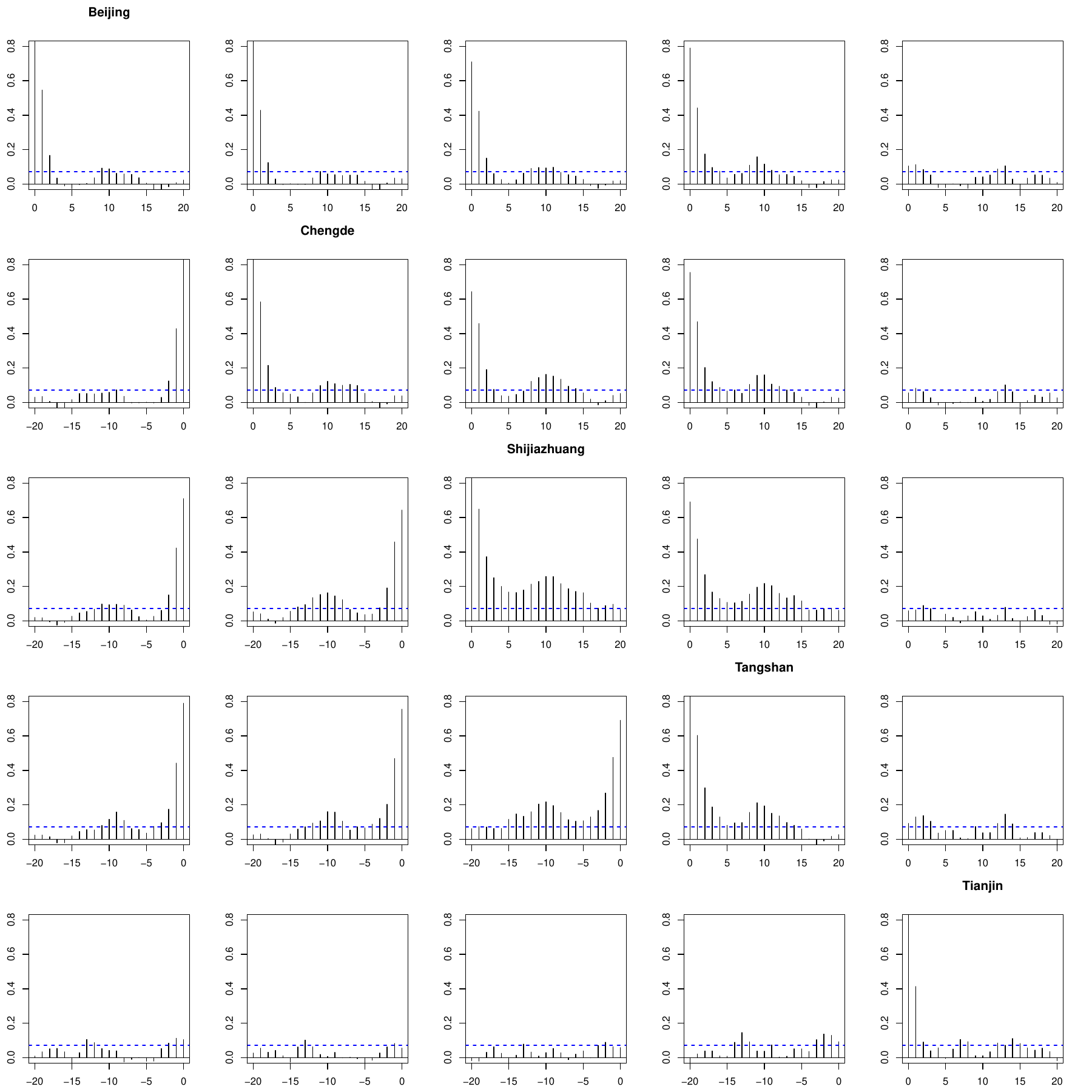}
\caption{Cross correlogram of logarithmic daily PM$_{10}$ readings at five randomly selected monitoring stations in Beijing, Tianjin and Hebei}
\label{figure:row}
\end{figure}

\begin{figure}[htbp]
\centering
\includegraphics[scale=0.5,page=1]{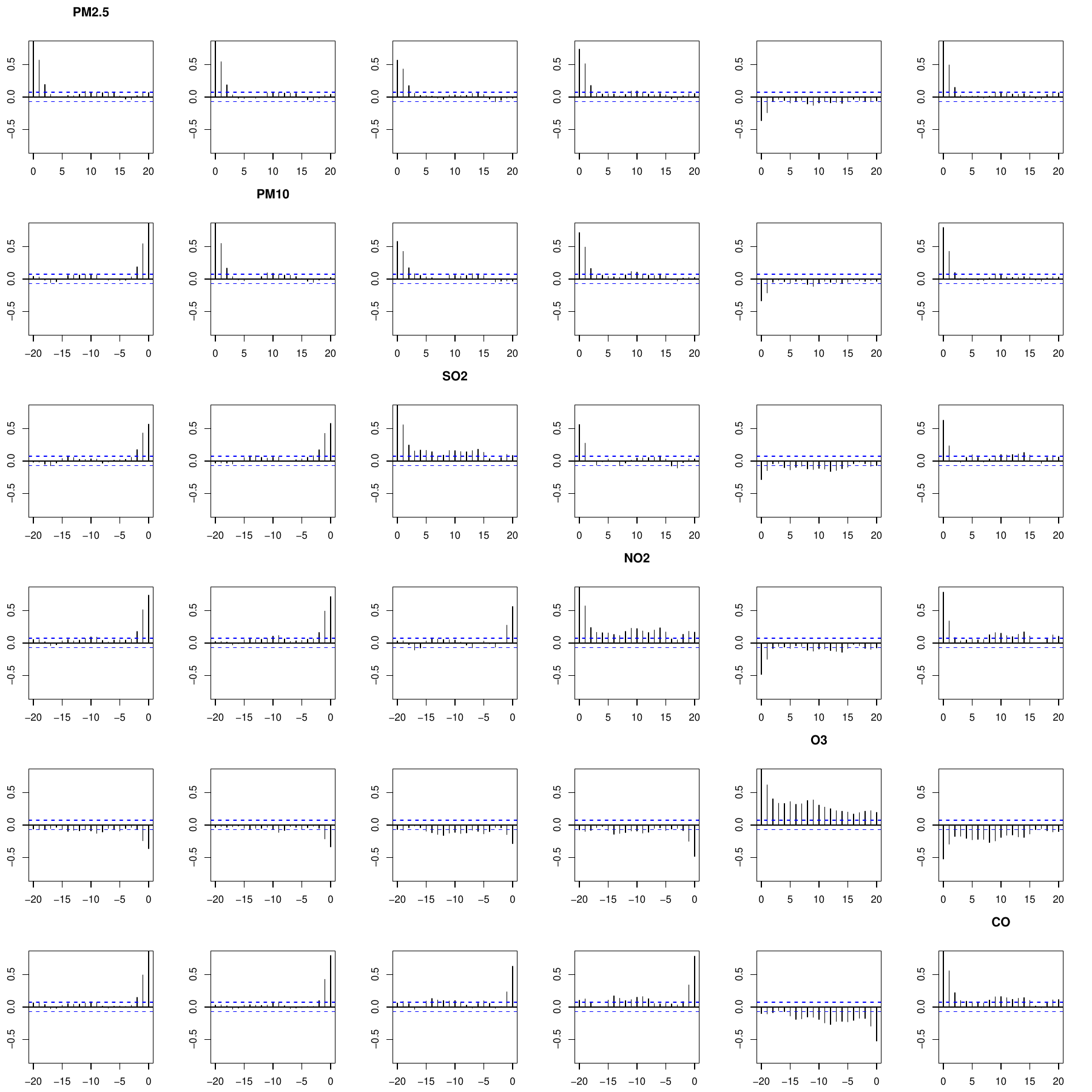}
\caption{Cross correlogram of logarithmic daily readings of all six pollutants for a station in Beijing}
\label{figure:col}
\end{figure}

In this appendix we provide additional figures of the air pollutant data used in the real data analysis. 
Figure \ref{figure:data} displays the time series plot of the $17\times 6=102$ series after removing the seasonal means. 
Figure \ref{figure:row} shows the cross correlogram of five randomly selected stations for PM$_{10}$, and Figure \ref{figure:col} displays the cross correlogram of all six pollutants for a station in Beijing.

\section{Proofs}\label{section:appendix}
%\subsection{Proofs of main theorems}

\begin{proof}[{\bf Proof of Proposition \ref{prop2}}]

In the following we provide the construction of $\bA_*=\bGamma^{(1)}\bUpsilon$, where the columns of $\bGamma^{(1)}$ is the $q$ orthonormal eigenvectors of $\bW^{(1)}$ and $\bUpsilon$ is a column permutation matrix. Let
\[
 Z_t^* = \bB\bU_t; \quad \bSigma_1^{(u)}=\E( Z_t^{*\top} Z_t^*); \quad
 Z_t= Z_t^*\big(\bSigma_1^{(u)}\big)^{-1/2}
\]
Clearly, $\bSigma_1^{(u)}$ is block
diagonal. Hence, the column of $ Z_t$ has same uncorrelated (segmentation) structure as that of $ Z_t^*$. 

We have
\begin{equation} \label{a4}
  \overrightarrow{ X_t} :=\bB\bU_t\bA^\top (\bSigma_1^{(x)})^{-1/2}
  = Z_t^* \bA^\top (\bSigma_1^{(x)})^{-1/2} = Z_t A_*^{\top}
\end{equation}
where $ A_*=\big(\bSigma_1^{(x)}\big)^{-1/2}\bA\big(\bSigma_1^{(u)}\big)^{1/2}$.
Note that
$\E( Z_t^\top Z_t)/p = \bI_{q} = \E(\overrightarrow{ X_t}^\top\overrightarrow{ X_t})/p$, 
where $\bI_{q}$ is the $q\times q$ identity matrix.
Hence $ A_*$ is orthonormal. Note that $B_*$ can be similarly defined.

Let the auto-cross-covariance matrices between the $i$-th and the $j$-th
rows of $\overrightarrow{ X_t}$ and the corresponding $ Z_t$ at lag $\tau$ be 
\bes
\bV^{(1)}_{\tau,i,j}
= \E\{\overrightarrow{ X}_{t+\tau}^\top \bE_{i,j}\overrightarrow{ X_t}\},
\mbox{\ \ and \ \ } 
\bV^{(z)}_{\tau,i,j} = \E\{\bZ_{t+\tau}^\top \bE_{i,j} Z_t\},
\ees
where $\bE_{i,j} = \be_i\be_j^\top \in \R^{p\times p}$ is the unit matrix with
1 at position $(i,j)$ and 0 elsewhere. Let
\begin{equation} \label{a9}
\bW^{(1)} = \sum_{\tau=-\tau_0}^{\tau_0} \sum_{i=1}^p\sum_{j=1}^p
\frac{\bV^{(1)}_{\tau,i,j}\big(\bV^{(1)}_{\tau,i,j}\big)^\top}{p^2},
\mbox{\ \ and \ \ } 
\bW^{(z)} = \sum_{\tau=-\tau_0}^{\tau_0} \sum_{i=1}^p\sum_{j=1}^p
\frac{\bV^{(z)}_{\tau,i,j}\big(\bV^{(z)}_{\tau,i,j}\big)^\top}{p^2}.
\end{equation}

Since $ Z_t$ has a column-wise uncorrelated structure,
$\bV^{(z)}_{\tau,i,j}$ is block diagonal for all $i,j$.
%$\bV^{(z,\tau,i,j)}  = diag(\bV^{(z,\tau,i,j)}_{1}, \ldots, \bV^{(z,\tau,i,j)}_{n_c})$
Hence, $\bW^{(z)}$ is block diagonal as well. Let
\begin{equation} \label{a8}
\bW^{(z)}
= \begin{pmatrix} \bW^{(z)}_{1} & 0 & \cdots & 0 \cr
0 & \bW^{(z)}_{2} & \cdots & 0 \cr \vdots & \vdots & \ddots & \vdots \cr
0 & 0 & \cdots & \bW^{(z)}_{n_c}\end{pmatrix}.
\end{equation}
Write the eigenvalue decomposition of each $\bW_i^{(z)}$ in \eqref{a8} as
\begin{equation} \label{a2}
\bW_i^{(z)} = \bGamma_i^{(z)} \bD_i \big(\bGamma_i^{(z)}\big)^\top,
\end{equation}
where $\bGamma_i^{(z)}$ is a $q_i\times q_i$ orthogonal matrix and $\bD_i$
is a diagonal matrix, consists of the eigenvalues of $\bW^{(z)}_i$ in descending
order $\lambda_{i1}\geq \lambda_{i2} \geq \ldots \geq \lambda_{iq_i}$. Hence
we can
write 
\begin{equation} \label{svd1}
\bW^{(z)} = \bGamma^{(z)} \bD \big(\bGamma^{(z)}\big)^\top, 
\end{equation}
where
\begin{equation} \label{a7}
\bGamma^{(z)} =\diag\big( \bGamma^{(z)}_1, \cdots, \bGamma^{(z)}_{n_c}\big), \quad
\quad \bD =\diag\big( \bD_1, \cdots, \bD_{n_c}\big),
\end{equation}
with $\bGamma^{(z)}$ orthogonal. 
Let the eigenvalue decomposition of $\bW^{(z)}$ (as a whole) be
\begin{equation} \label{svd2}
\bW^{(z)} = \bGamma^{(z*)} \bD^* \big(\bGamma^{(z*)}\big)^\top,
\end{equation}
where the eigenvalues in the diagonal matrix $\bD^*$ is in descending order.
Comparing \eqref{svd1} and \eqref{svd2}, these exists a column
permutation matrix $\bUpsilon$ such that
$\bGamma^{(z*)}=\bGamma^{(z)}\bUpsilon^{\top}$, and
$\bD^*=\bUpsilon\bD\bUpsilon^\top$.

Since $\overrightarrow{ X_t} =  Z_t  A_*^\top$ in \eqref{a4}, we have
$\bV^{(1)}_{\tau,i,j} =  A_* \bV^{(z)}_{\tau,i,j}  A_*^\top$, Hence
\[
\bW^{(1)} = A_* \bW^{(z)}  A_*^\top
= A_*\bGamma^{(z)} \bD(\bGamma^{(z)})^\top A_*^\top
= A_*\bGamma^{(z*)}\bD^* A_*(\bGamma^{(z*)})^\top A_*^\top
=\bGamma^{(1)}\bD^*\big(\bGamma^{(1)}\big)^\top
\]
where $\bGamma^{(1)} =  A_* \bGamma^{(z*)}$. Since
$\bGamma^{(1)}$ is an orthogonal matrix and $\bD^*$ is a diagonal matrix
with descending diagonals, it is clear that $\bD^*$ contains the eigenvalues
of $\bW^{(1)}$ and $\bGamma^{(1)}$ contains the corresponding eigenvectors. 

Now it follows from (\ref{a4}) that
\begin{equation} \label{a10}
\overrightarrow{ X_t}\bGamma^{(1)} =  Z_t  A_*^\top  A_* \bGamma^{(z*)}
=  Z_t \bGamma^{(z*)}= Z_t \bGamma^{(z)}\bUpsilon^\top .
\end{equation}
Or
\begin{equation} \label{a11}
\overrightarrow{ X_t}\bGamma^{(1)}\bUpsilon = Z_t \bGamma^{(z)}.
\end{equation}

Since the columns of $ Z_t \bGamma^{(z)}=\overrightarrow{ X_t}\bGamma^{(1)}\bUpsilon$
has the same uncorrelated structure as that of $ Z_t$,
%Therefore
%$\overrightarrow{ X_t}\bGamma^{(1)}\bUpsilon^{-1}
%= X_t (\bSigma^{(1)}_1)^{-1/2} \bGamma^{(1)}\bUpsilon^{-1}$ has the desired
%partition structure.
$\bGamma^{(1)}\bUpsilon$ is the desired orthogonal transformation $\bA_*$
we are looking for. 
\end{proof}

As $|\P(\cE_1\cap \cE_2) - \P(\cE_1)\P(\cE_2)|\le 1/4 \le e^{-1}$ for all event pairs $\{\cE_1,\cE_2\}$, 
\bes
&& \sup_{t}\Big\{\Big|\P(\cE_1\cap \cE_2) - \P(\cE_1)\P(\cE_2)\Big|:
\cE_1\in \sigma( X_{s+\tau}\otimes  X_s, s\le t,\tau\le\tau_0),
\cE_2\in \sigma( X_{s+\tau}\otimes  X_s, s\ge t+k)\Big\}
\cr &\le & \sup_{t}\Big\{\Big|\P(\cE_1\cap \cE_2) - \P(\cE_1)\P(\cE_2)\Big|:
\cE_1\in \sigma( X_s, s\le t+\tau_0),
\cE_2\in \sigma( X_s, s\ge t+k)\Big\}
\cr &\le & \exp\Big[- \max\{1,c_0(k-\tau_0)^{r_1}\}\Big]
\cr &\le & \exp\Big[ - c_0(1+c_0^{1/r_1}\tau_0)^{-r_1}k^{r_1}\Big],
\ees
where $\otimes$ denotes the Kronecker product for matrices.
Thus, the quadratic process $( X_{t+\tau}\otimes  X_t, 0\le \tau\le\tau_0)$
also satisfies the $\alpha$-mixing condition with
$c_0$ in Assumption \ref{asmp:mixing} replaced by the smaller $c_0(1+c_0^{1/r_1}\tau_0)^{-r_1}$.

\begin{proof}[{\bf Proof of Theorem \ref{th-1new}.}]
As $\| \bV\|_{\rm op}\le q\|\bV\|_{\max}$ for all matrices $\bV\in \R^{q\times q}$, it suffices
to bound the estimation error under the $\|\cdot\|_{\max}$ norm. In the proof, we consider a more general case that $\E X_t$ is not necessary 0 and let
\begin{equation*} 
\wh\Sigma^{(1)}_{\tau,i,j}
= \sum_{t=1}^{T-\tau}
{({ X}_{t+\tau}-\bmuhat_{\tau,+})^\top \bE_{i,j}( X_t-\bmuhat_{\tau,-})
\over T-\tau}
, \;\;
 \wh\bSigma_1^{(x)}=\frac1p \sum_{i=1}^p \wh\Sigma^{(1)}_{0,i,i},
\end{equation*}
and
\bes
\bmuhat_{\tau,+} = \frac{\sum_{t=1}^{T-\tau} X_{t+\tau}}{T-\tau},\;\;
\bmuhat_{\tau,-} = \frac{\sum_{t=1}^{T-\tau} X_{t}}{T-\tau},\;\;
\bSigmatil^{(1)}_{\tau,i,j}
= \sum_{t=1}^{T-\tau}\frac{ X_{t+\tau} \bE_{i,j} X_t}{T-\tau}.
\ees

Note that $\etil_{k}^\top (\bSigmatil^{(1)}_{\tau,i,j} - \bSigma^{(1)}_{\tau,i,j})\etil_l = X_{t+\tau,i,k}X_{t,j,l}-\E X_{t+\tau,i,k}X_{t,j,l}$, where $\etil_k$ is a $q\times1$ vector with 1 at $k$th element and 0 otherwise. Pick $a>0$ and $u>\E X_{t+\tau,i,k}X_{t,j,l}$, then
\begin{align*}
&\P \left(|X_{t+\tau,i,k}X_{t,j,l}-\E X_{t+\tau,i,k}X_{t,j,l}| >u \right)\\
=&\P \left(|X_{t+\tau,i,k}X_{t,j,l}-\E X_{t+\tau,i,k}X_{t,j,l}| >u,  |X_{t+\tau,i,k}| >u^a\right) \\
&\quad + \P \left(|X_{t+\tau,i,k}X_{t,j,l}-\E X_{t+\tau,i,k}X_{t,j,l}| >u,  |X_{t+\tau,i,k}| \le u^a\right) \\
\le&\P \left( |X_{t+\tau,i,k}| >u^a\right) + \P \left(|X_{t,j,l}| >u^{1-a}\right) \\
\le& c_1\exp(-c_2u^{ar_2})+c_1\exp(-c_2u^{(1-a)r_2}) \\
\le& 2c_1\exp(-c_2u^{r_2/2}).
\end{align*}
Let $1/\beta_1=1/r_1+2/r_2$ and $1/\beta_2=1/r_1+1/r_2$. Hence, by Theorem 1 of \cite{merlevede2011bernstein}, Bernstein inequality for $\alpha$ strong mixing processes, we have 
\begin{align*}
\P\left(|\etil_{k}^\top (\bSigmatil^{(1)}_{\tau,i,j} - \bSigma^{(1)}_{\tau,i,j})\etil_l| \ge x\right) &\le T\exp\left(-C_1T^{\beta_1}x^{\beta_1} \right)+\exp\left(-C_2 Tx^2 \right),\\
\P\left( |\etil_k^\top \bmuhat_{\tau,+}^\top E_{ij} \bmuhat_{\tau,-}\etil_l |\ge x \right) &\le T\exp\left(-C_1T^{\beta_2}x^{\beta_2/2} \right)+\exp\left(-C_2 Tx \right).
\end{align*}
Then,
\begin{align*}
\P\left(|\etil_{k}^\top (\bSigmahat^{(1)}_{\tau,i,j} - \bSigma^{(1)}_{\tau,i,j})\etil_l| \ge x\right) \le& T\exp\left(-CT^{\beta_1}x^{\beta_1} \right)+T\exp\left(-CT^{\beta_2}x^{\beta_2/2} \right)\\
&+\exp\left(-C Tx^2 \right) +\exp\left(-C Tx \right).
\end{align*}
With probability at least $1-\epsilon_T$, for any $1\le i,j\le p$, we have
\begin{align}
\left\| \bSigmahat^{(1)}_{\tau,i,j} - \bSigma^{(1)}_{\tau,i,j} \right\|_{\max}\le C_1'\sqrt{\frac{\log(pq/\epsilon_T)}{T}} + C_1' \frac{[\log(Tpq/\epsilon_T)]^{1/\beta_1}} {T} + C_1'\frac{[\log(Tpq/\epsilon_T)]^{2/\beta_2}} {T^2}.
\end{align}
Then \eqref{th-1-1new} follows. The proof is complete as \eqref{th-1b-1new} follows from Wedin's theorem, \eqref{th-1-1new}, \eqref{est-1}, and \eqref{est-2}. 
%and \eqref{th-1-3new} follows from Wedin's theorem, Proposition \ref{prop2}, Lemma \ref{th-2}, \eqref{th-1b-1new} directly.
By Wedin's theorem, \eqref{th-1b-1new} implied directly that $||\hat{\Gamma}^{(1)}-\Gamma^{(1)}||\leq C_2\eta_{T,p,q}$. Hence \eqref{th-1-3new} 
holds by Proposition \ref{prop2} and Lemma \ref{th-2}.
\end{proof}

\begin{proof}[\bf Proof of Theorem \ref{th-1-poly}.]
From Assumption \ref{asmp:tail2}, it yields
\begin{align*}
&\max_t\max_{i,j,k,l}\P \left(|X_{t_\tau,i,k}X_{t,j,l}-\E X_{t_\tau,i,k}X_{t,j,l}| >u \right) \le c_1' x^{-r_1}.
\end{align*}
By the Fuk-Nagaev type inequality stated in Theorem 6.2, or equation (6.19a), of \cite{rlo2000theorie},
\begin{align*}
\P\left(|\etil_{k}^\top (\bSigmatil^{(1)}_{\tau,i,j} - \bSigma^{(1)}_{\tau,i,j} )\etil_l| \ge x\right) &\le 4\exp(-C_1Tx^2)+4C_2 T(Tx)^{-\frac{(r_1+1)r_2}{r_1+r_2}} ,\\
\P\left( |\etil_k^\top \bmuhat_{\tau,+}^\top E_{ij} \bmuhat_{\tau,-}\etil_l |\ge x \right) &\le 4\exp(-C_1Tx^2)+4C_2 T(Tx)^{-\frac{2(r_1+1)r_2}{r_1+2r_2}}.
\end{align*}
Then,
\begin{align*}
\P\left(|\etil_{k}^\top (\bSigmahat^{(1)}_{\tau,i,j} - \bSigma^{(1)}_{\tau,i,j})\etil_l| \ge x\right) \le& C T(Tx)^{-\frac{(r_1+1)r_2}{r_1+r_2}}+8\exp(-C_1Tx^2).
\end{align*}
Let $\beta_3=r_2(r_1+1)/(r_1+r_2)>1$. With probability at least $1-\epsilon_T$, for any $1\le i,j\le p$, we have
\begin{align}
\left\| \bSigmahat^{(1)}_{\tau,i,j} - \bSigma^{(1)}_{\tau,i,j} \right\|_{\max}\le C_1' \frac{(Tpq/\epsilon_T)^{1/\beta_3}}{T} +  C_1'\sqrt{\frac{\log(pq/\epsilon_T)}{T}}.
\end{align}
The rest is the same as the proofs in Theorem \ref{th-1new}.
\end{proof}

For any group $G\subset \{1,\ldots,q\}$, define the eigengap as
\bel{eq:deltag}
\Delta_G = \min\Big\{|\lam_j-\lam_k|: j\in G, k\not\in G\Big\}.
\eel
Let $\bGamma_{G}^{(1)}=(\bgamma_j)_{j\in G}$. We say that the group $G$ is {\it identifiable} if $\Delta_G>0$ as the column space of $\bGamma^{(1)}_G$ would then be the span of $|G|$ eigenvectors for all versions of $\bGamma^{(1)}$.

\begin{lemma}\label{th-2}
Suppose conditions of Theorem \ref{th-1new} (resp. Theorem \ref{th-1-poly}) hold. For a certain $G\subset \{1,\ldots,q\}$, assume $C_1\eta_{T,p,q} < \Delta_G/2$ with the $\eta_{T,p,q}$ in (\ref{th-1-2new}) and $C_1$ given in Theorem \ref{th-1new} (resp. Theorem \ref{th-1-poly}) depending on $c_*, c^*,r_1,r_2$ only.
Let $\bP_{G} = \bGamma^{(1)}_{G}\Big(\bGamma^{(1)}_{G}\Big)^\top$ be the orthogonal projection to the corresponding eigenspace, and $\bPhat_G = \bGammahat^{(1)}_G\Big(\bGammahat^{(1)}_G\Big)^\top$ be the sample version. Then, in the event $\Omega_T$ given in Theorem \ref{th-1new} (resp. Theorem \ref{th-1-poly}) with probability at least $1-\epsilon_T$, by \eqref{th-1b-1new},
\bel{th-2-1}
\Big\| \bPhat_G - \bP_G\Big\|_{\rm op} \le  \frac{C_1\eta_{T,p,q}}{\Delta_G - C_1 \eta_{T,p,q}} < 1 .
%\hbox{with }\
%\eta^{**}_{T,p,G} = \frac{\eta^{**}_{T,p,q}}{\Delta_G - \eta^{**}_{T,p,q}} < 1.
\eel
% In particular, if $2\eta^{**}_{T,p,q} \le \lam_j-\lam_{j+1}$, then in the event $\Omega_T$
% \bes
% \Big\| \bPhat_{\{1,\ldots,j\}} - \bP_{\{1,\ldots,j\}}\Big\|_{\rm op}
% \le \frac{\eta^{**}_{T,p,q}}{\lam_j-\lam_{j+1} - \eta^{**}_{T,p,q}} < 1.
% \ees
\end{lemma}

% Write the eigenvalue decomposition of $\bW^{(1)}$ as
% \bes
% \bW^{(1)} = \bGamma^{(1)}\bD\big(\bGamma^{(1)}\big)^\top,
% \ees
% where $\bD=\diag(\lam_1,\ldots,\lam_q)$ contains  the eigenvalues $\lam_1\ge \cdots \ge \lam_q$ and $\bGamma^{(1)} = (\bgamma_1,\ldots,\bgamma_q)$ contains the corresponding eigenvectors. 

By the structure assumption in (\ref{a1}), there exists a group structure $G_c=\{ G_1,...,G_{n_c}\}$, such that the columns of different groups
\bel{gamma-decomp1}
X_t\Big(\bSigma^{(1)}_1\Big)^{-1/2}\bGamma^{(1)}_{G_i},\ i = 1,\ldots,n_c,
\eel
are uncorrelated across all time lags. Let $G$ be one of the $G_i$ or a union of several $G_i$. It follows from Lemma \ref{th-2} that when the group $G$ is identifiable with a sufficiently large eigengap $\Delta_G$, $\cM(\widehat A_{*G})$ is a consistent estimator of the subspace $\cM(A_{*G})$, even in the presence of tied eigenvalues within $G$ or within $G^c$.

% However, the representation (\ref{gamma-decomp1}) is not unique. An alternative representation
% \bel{alt-decomp1}
%  X_t\Big(\bSigma^{(x)}_1\Big)^{-1/2}\bPi_{H_i}, \ i=1,\ldots, n_c,
% \eel
% based on an orthonormal matrix $\bPi=(\wt\bgamma_1,\ldots,\wt\bgamma_q)$ and a partition $\{H_1,\ldots,H_{n_c}\}$
% of $\{1,\ldots,q\}$ is equivalent to (\ref{gamma-decomp1}) if and only if
% \bes
% \bPi_{H_i}\bPi_{H_i}^\top  = \bGamma^{(1)}_{G_i}\big(\bGamma^{(1)}_{G_i}\big)^\top,\ i=1,\ldots, n_c.
% \ees
% This includes rotation of the basis vectors within the group and also permutation of indices of the basis vectors. 

\begin{proof}[\bf Proof of Theorem \ref{th-3}.] Let $U\diag(\phi_1,\ldots,\phi_r)\wh U^\top$ be the SVD of
$\bP_{G_{i_0}}\bPhat_{G_{i_0}}$ with $\phi_1\ge\cdots\ge\phi_r$ and
$n_{i_0}=|G_{i_0}|$, $1\le i_0\le n_c$. Define
$\bPi_{G_{i_0}} = U\wh U^\top \bGammahat^{(1)}_{G_{i_0}}$.
As $\wh U\wh U^\top = \bPhat_{G_{i_0}}
= \bGammahat^{(1)}_{G_{i_0}}\big(\bGammahat^{(1)}_{G_{i_0}}\big)^\top$,
$\bPi_{G_{i_0}}$ has orthonormal columns. Note that $UU^\top = \bGamma^{(1)}_{G_{i_0}}\big(\bGamma^{(1)}_{G_{i_0}}\big)^\top$. Moreover,
\bes
\|\bPi_{G_{i_0}} - \bGammahat^{(1)}_{G_{i_0}}\|_{\rm op}
= \|(U-\wh U)\wh U^\top \bGammahat^{(1)}_{G_{i_0}}\|_{\rm op}
\le \|U-\wh U\|_{\rm op}
= \max_{1\le j\le n_{i_0}} \|(U-\wh U)\be_j\|_2.
\ees
Let $u_j=U\be_j$ and $\wh u_j = \wh U \be_j$, where $e_j$ is a $q\times 1$ vector with 1 at $j$-th element and 0 otherwise. As $$(\bP_{G_{i_0}} - \bPhat_{G_{i_0}})(u_j,\wh u_j)=(u_j-\wh U\wh U^\top U e_j, UU^\top \wh U e_j-\wh u_j)=(u_j-\phi_j\wh u_j, \phi_j u_j-\wh u_j), $$
the span of the columns $u_j$ and $\wh u_j$ forms invariance subspaces of $\bP_{G_{i_0}} - \bPhat_{G_{i_0}}$. Then a linear combination of $u_j$ and $\wh u_j$ forms eigenvectors of $\bP_{G_{i_0}} - \bPhat_{G_{i_0}}$. Elementary calculation shows that $\bP_{G_{i_0}} - \bPhat_{G_{i_0}}$ has eigenvectors $u_j - \phi_j^{-1}\big(1\pm \sqrt{1-\phi_j^2}\big)\wh u_j$.
As
\bes
0 =\Big\langle u_j - \phi_j^{-1}\big(1 + \sqrt{1-\phi_j^2}\big)\wh u_j,
u_j - \phi_j^{-1}\big(1 - \sqrt{1-\phi_j^2}\big)\wh u_j\Big\rangle
= 1 - 2 u_j^\top\wh u_j/\phi_j + 1,
\ees
$\|(U-\wh U)\be_j\|_2^2 = 2(1-\phi_j)$, so that by Lemma \ref{th-2}, if $2C_1\eta_{T,p,q}<\Delta$, then
\bes
\|\bPi_{G_{i_0}} - \bGammahat^{(1)}_{G_{i_0}}\|_{\rm op}^2
= 2(1-\phi_r) \le 2\Big(1 - \sqrt{1-\|\bP_{G_{i_0}} - \bPhat_{G_{i_0}}\|_{\rm op}^2}\Big)
\le 8C_1^2\eta_{T,p,q}^2/\Delta^2.
\ees
For this potentially random $\bPi_{G_{i_0}}$, the inseparability condition provides a $\rho^*$-connection graph $(G_{i_0}, E_{i_0})$ with $H_{i_0}=G_{i_0}$.
As the eigenvalues of $\bV^{(1)}_{0,j,j}$ are within $[c_*/c^*,c^*/c_{*}]$,
for $(k,\ell)\in E_{i_0}$ and in an event $\Omega_T$ with probability at least $1-\epsilon_T$,
\bes
&\left| \big\langle\bgammahat_k, \bVhat^{(1)}_{0,i,i} \bgammahat_k\big\rangle - \big\langle\bgammahat_k, \bV^{(1)}_{0,i,i} \bgammahat_k\big\rangle \right| \le C_1\eta_{T,p,q} \le (C_1 c^*\eta_{T,p,q}/c_{*}) \big\langle\bgammahat_k, \bV^{(1)}_{0,i,i} \bgammahat_k\big\rangle,\\
&\left| \big\langle\bgammahat_k, \bV^{(1)}_{0,i,i} \bgammahat_k\big\rangle - \big\langle\bgamma_k, \bV^{(1)}_{0,i,i} \bgamma_k\big\rangle \right| \le 2(c^{*}/c_{*}) \|\bgammahat_k -\bgamma_k \|_{\rm op} \le 4\sqrt{2}(C_1 c^{*}/c_{*}) \eta_{T,p,q}/\Delta,
\ees
in view of $\|\bgammahat^{(1)}_k -\bgamma_k \|_{\rm op} \le \|\bPi_{G_{i_0}} - \bGammahat^{(1)}_{G_{i_0}}\|_{\rm op} \le 2\sqrt{2}C_1 \eta_{T,p,q}/\Delta$. 
Thus, for any $1\le i_0\le n_c$ and $k,\ell\in G_{i_0}$,
\bes
\hrho^{(1)}_{k,\ell}
&\ge& \frac{1}{1+ C_1c^*\eta_{T,p,q}/c_*}\max_{i,j,\tau}
\frac{\big|\big\langle \bgammahat_k, \bVhat^{(1)}_{\tau,i,j} \bgammahat_\ell\big\rangle\big|}
{\big\{\big\langle\bgammahat_k, \bV^{(1)}_{0,i,i} \bgammahat_k\big\rangle
         \big\langle \bgammahat_\ell,\bV^{(1)}_{0,j,j} \bgammahat_\ell\big\rangle \big\}^{1/2}} \cr 
&\ge& \frac{(1 + 4\sqrt{2}C_1 c^{*2}\eta_{T,p,q}/(c_{*}^2\Delta))^{-1}}
{1 + C_1c^*\eta_{T,p,q}/c_*}\max_{i,j,\tau}
\frac{\big|\big\langle \bgammahat_k, \bVhat^{(1)}_{\tau,i,j} \bgammahat_\ell\big\rangle\big|}
{\big\{\big\langle\bgamma_k, \bV^{(1)}_{0,i,i} \bgamma_k\big\rangle
         \big\langle \bgamma_\ell,V^{(1)}_{0,j,j} \bgamma_\ell\big\rangle \big\}^{1/2}} \cr 
&\ge& \frac{\rho^* - \big(C_1 + 4\sqrt{2}C_1 c^*/(c_{*}\Delta)\big)\eta_{T,p,q}}
{(1 + 4\sqrt{2}C_1 c^{*2}\eta_{T,p,q}/(c_{*}^2\Delta))(1 + C_1c^*\eta_{T,p,q}/c_*)} \cr
&\ge& 0.9 \rho^* - C_3 \eta_{T,p,q} \cr
&=:& \rho_+.
\ees
Note that the last inequality follows from the first part of condition \eqref{cond}, by setting $80\sqrt{2}C_1 (c^*/c_*)^2 \eta_{T,p,q}<\Delta$ and $20C_1(c^*/c_*)(c^*-c_*)\eta_{T,p,q}<\Delta$.
Similarly, for any $1\le i_0\le n_c$ and $k\in G_{i_0}$ and $\ell\not\in G_{i_0}$,
\begin{align*}
\hrho^{(1)}_{k,\ell} \le \frac{ \big(C_1 + 4\sqrt{2}C_1 c^*/(c_{*}\Delta)\big)\eta_{T,p,q} }
{(1 - 4\sqrt{2}C_1 c^{*2}\eta_{T,p,q}/(c_{*}^2\Delta))(1 - C_1c^*\eta_{T,p,q}/c_*)} =C_4 \eta_{T,p,q} =: \rho_-    .
\end{align*}
Note that $C_3\le (C_1 + 4\sqrt{2}C_1 c^*/(c_{*}\Delta))/1.05^2$ and $C_4\le (C_1 + 4\sqrt{2}C_1 c^*/(c_{*}\Delta))/0.95^2$. By the second part of condition \eqref{cond}, with $\kappa_2=1.01\cdot (C_1 + 4\sqrt{2}C_1 c^*/(c_{*}\Delta))$, it holds that $\rho_+ >\rho_-$.

For $\wh r$ defined in \eqref{a14mod}, we have
\begin{align*}
\max_{1\le j< r}\frac{\wh \rho_{(j)}+\delta_T} {\wh\rho_{(j+1)}+\delta_T}  \le \frac{1+\delta_T}{\rho_+ +\delta_T} ,\quad  \frac{\wh \rho_{(r)}+\delta_T} {\wh\rho_{(r+1)}+\delta_T} > \frac{\rho_+ +\delta_T}{\delta_T},\quad   \max_{r< j\le q_0}\frac{\wh \rho_{(j)}+\delta_T} {\wh\rho_{(j+1)}+\delta_T}  \le \frac{\rho_- +\delta_T}{\delta_T} .
\end{align*}
By $\delta_T\le (\rho^*)^2/2$ and the second part of condition \eqref{cond}, under certain $\kappa_2$, we have $\delta_T \le \rho_+^2 +2\rho_+ \delta_T$. It follows that $\wh r=r$ and $\wh E=E$. That is, in the event $\Omega_T$, every $G_{i_0}$ is connected through $E_{i_0}$ and separated from $G_{i_0}^c$. Therefore, $\wh n_c=n_c$ and $\wh G_i=G_i, 1\le i\le n_c,$ up to some permutation of group indices. 

In our analysis, we set $\kappa_1=\max\{80\sqrt{2}C_1 (c^*/c_*)^2, 20C_1(c^*/c_*)(c^*-c_*)\}$ and $\kappa_2=1.01\cdot (C_1 + 4\sqrt{2}C_1 c^*/(c_{*}\Delta))$. But other specification of $\kappa_1,\kappa_2$ may also work.
\end{proof}

\end{document}